\documentclass[]{aa}
\makeatletter
\renewcommand*\aa@pageof{, page \thepage{} of \pageref*{LastPage}}
\makeatother

\newcommand\xrt{{\it Swift-{\it XRT}}}

\newcommand\cha{\textit{Chandra}}
\newcommand\xmm{{XMM-{\it Newton}}}
\newcommand\NuSTAR{{\it NuSTAR}}

\newcommand\lat{{\it Fermi}-LAT}
\def\flu{{erg\,s$^{-1}$\,cm$^{-2}$}}

\usepackage{natbib}
\usepackage{float}
\usepackage{color}
\usepackage{graphicx}
\usepackage{textcomp,gensymb}
\usepackage{array,amsmath}
\usepackage{mathtools}
\usepackage{multirow,makecell}
\usepackage{enumitem}
\usepackage{longtable}
\usepackage{hyperref}
\begin{document} 

    \title{A new look at the extragalactic Very High Energy sky: searching for TeV-emitting candidates among the X-ray bright, non-Fermi detected blazar population}

    \author{Stefano~Marchesi\inst{1,2,3}, Antonio~Iuliano\inst{4}, Elisa~Prandini\inst{5}, Paolo~Da Vela\inst{2}, Michele~Doro\inst{5}, Serena~Loporchio\inst{6,7}, Davide~Miceli\inst{5}, Chiara~Righi\inst{8}, Roberta~Zanin\inst{9}, Ettore~Bronzini\inst{1,2}, Cristian~Vignali\inst{1,2}
    }
    \institute{Dipartimento di Fisica e Astronomia (DIFA) Augusto Righi, Università di Bologna, via Gobetti 93/2, I-40129 Bologna, Italy
    \and INAF-Osservatorio di Astrofisica e Scienza dello Spazio (OAS), via Gobetti 93/3, I-40129 Bologna, Italy
    \and Department of Physics and Astronomy, Clemson University, Kinard Lab of Physics, Clemson, SC 29634, USA
    \and Sezione INFN di Napoli, Complesso universitario di Monte S. Angelo ed. 6 via Cintia, 80126, Napoli, Italy
    \and INFN Sezione di Padova and Università degli Studi di Padova, Via Marzolo 8, I-35131 Padova, Italy
    \and Dipartimento Interateneo di Fisica, Politecnico di Bari, via Amendola 173, 70125 Bari
    \and Sezione INFN di Bari, via Orabona 4, 70125, Bari
    \and INAF – Osservatorio Astronomico di Brera, Via E. Bianchi 46, 23807 Merate, Italy
    \and Cherenkov Telescope Array Observatory gGmbH, Via Piero Gobetti, 93/3, 40129, Bologna, Italy
               }

\titlerunning{A new look at the extragalactic Very High Energy sky}
\authorrunning{Marchesi et al.}

\abstract
{We present the results of a multi-wavelength study of blazars selected from the 5th ROMABZCAT catalog. We selected from this sample a subsample of 2435 objects having at least one counterpart in one of the three main archival X-ray catalogs, which is, the fourth release of the \xmm\ Survey Science Catalogue, the second release of the \cha\ Source Catalog, and the second \textit{Swift} X-ray Point Source catalog of detections by \xrt, or in the recently released eROSITA--DE Data Release 1 catalog. 
We first searched for different multi-wavelength trends between sources with a $\gamma$--ray counterpart in the \lat\ 14-year Source Catalog (4FGL--DR4) and sources lacking one. We find that the non--4FGL sources are on average fainter both in the X-rays and in the radio with respect to the 4FGL--detected ones, but the two samples have similar X-ray--to--radio flux ratios, as well as synchrotron peak frequencies.
We then focused on the 1007 non--$\gamma$--ray detected population, to determine if there is a sample of X-ray sources that could be TeV emitters. We find that a large number of sources, mostly BL Lacs or BL Lacs with host-galaxy contribution to the spectral energy distribution, have large synchrotron peak frequency and X-ray to radio flux ratio, two properties that characterize the vast majority of known TeV emitters. With respect to these known TeV emitters, our targets have X-ray fluxes $\sim$1 order of magnitude fainter.
We then computed the 0.2--12\,keV and 20\,GeV -- 300\,TeV fluxes for the known 5BZCAT TeV emitters, and determined the existence of a direct correlation between X-ray and TeV fluxes in the BL Lacs population. 
We used this trend to estimate the VHE flux of our targets, and found a promising sample of sources for follow--up observations with current or future, more sensitive, Cherenkov telescopes, first and foremost the Cherenkov Telescope Array Observatory.
}

\keywords{BL Lacertae objects: general; Galaxies: active; Gamma rays: galaxies; X-rays: galaxies.}

\maketitle

\section{Introduction}\label{sec:intro}
Blazars are accreting supermassive black holes (SMBHs), or active galactic nuclei (AGN), whose relativistic jets are pointed in the direction of the observer. This causes the so-called ``Doppler boosting'', a significant enhancement of the source luminosity due to the relativistic speed of the particles causing the emission. The spectral energy distribution (SED) of blazars is characterized by two clear bumps  \citep[e.g.,][]{abdo10}: the first one at lower frequencies/energies, the so-called ``synchrotron peak'', where the synchrotron emission is caused by the relativistic electrons in the jets. The second one at higher frequencies/energies, the ``inverse Compton'' one, is instead caused by the interaction and subsequent up-scattering in frequency of the synchrotron--produced photons with the same relativistic electrons \citep[in the so--called synchrotron self-Compton scenario; e.g.,][]{kirk98}.

The frequency of these two peaks can vary significantly depending on several parameters, most prominently the blazars luminosity and their class \citep[e.g.,][]{padovani95,ghisellini98,giommi99, prandini22}. For example, the synchrotron peak can be found at frequencies as low as $\nu_{\rm synch}$ = 10$^{12}$\,Hz, which is, in the far infrared band \citep[e.g.,][]{chen09}, and as high as  $\nu_{\rm synch}$ = 10$^{18}$\,Hz, in the X-ray band \citep[which is, conventionally, between 0.1 and several hundreds keV; see, e.g.,][]{chang19}.

The X-ray band, in particular, is a key region of the blazars SED. In the most luminous and powerful blazars, most of which are detected at redshifts $z>$1, X-ray observations are key to detect and constrain the shape of the inverse Compton bump rising part  \citep[e.g.,][]{tavecchio00,sbarrato15,marcotulli17,ghisellini19,marcotulli20}. In a specific class of sources, aptly called High-synchrotron peaked (HSPs; also defined as HBLs) blazars \citep[e.g.,][]{massaro11a,arsioli15,chang17,chang19}, the X-ray band is the region where the synchrotron peak is located, as mentioned above. In these extreme objects, the observational evidence supports a scenario where the X-ray emission is correlated with the Very High Energy (VHE; conventionally, this means that photons with energies larger than 50\,GeV have been detected) one. For example, it has been shown \citep{massaro11a,massaro11b} that known VHE emitters tend to lie, preferentially, in a well defined region of the synchrotron peak energy versus X-ray spectral curvature around the peak \citep[as defined, e.g., in][]{tramacere07,massaro08,tramacere11}.

As a consequence of this correlation, previous works have been focused on using the X-ray emission of blazars to select promising VHE emitters. This is often done by combining the X-ray information with the one from at least another band \citep[such as the infrared one, see, e.g.,][or the radio one, see, e.g., \citealt{bonnoli15}]{massaro13, giommi24}. So far, however, the overwhelming majority of the sources analyzed in these works were known $\gamma$--ray emitters, and in particular sources detected by the \textit{Fermi} Large Area Telescope (LAT). Indeed, predictions for detections of extragalactic sources in VHE surveys fairly often use as a starting point extrapolations based on \textit{Fermi}--LAT luminosity functions in the MeV to GeV band. While this is certainly a tested and reliable method to estimate VHE fluxes, it could nonetheless miss a population of sources that might be not detected in the $\gamma$--ray band and still be visible at larger energies. Within this framework, in this paper we will perform a multi--wavelength study of a broad sample of radio--selected blazars, with a specific focus on a sample of sources that lack \lat\ data while at the same time being detected in the X-rays.

The work is organized as follows: in Section~\ref{sec:sample} we present the sample used in this work and the different multi-wavelength catalogs used in the analysis. In Section~\ref{sec:multiwave}, we analyze the multi-wavelength properties of our targets, and in particular we compare those sources with a \lat\ counterpart with those without one. In Section~\ref{sec:TeV_predictions} we then use the multi--wavelength properties of the non--\lat\ detected sources, and in particular their X-ray flux, to make predictions on their TeV emission. We finally summarize the main results of this work and discuss possible future developments in Section~\ref{sec:conclusions}. Through the rest of the work, we assume a flat $\Lambda$CDM cosmology with H$_0$=69.6\,km\,s$^{-1}$\,Mpc$^{-1}$, $\Omega_m$=0.29 and $\Omega_\Lambda$=0.71 \citep{bennett14}. Errors are quoted at the 90\,\% confidence level, unless otherwise stated.

\section{Sample selection}\label{sec:sample}

All the sources analyzed in this work are initially selected from the 5th ROMABZCAT -- Roma--BZCAT Multi--Frequency Catalog of Blazars \citep[hereafter 5BZCAT,][]{massaro15}. This catalog contains coordinates and multi-frequency data of 3561 sources that are either confirmed blazars or have multi--wavelength properties that strongly support a blazar nature. All the objects in the sample have a radio band detection. As reported in the description of the 5BZCAT catalog (\url{https://heasarc.gsfc.nasa.gov/w3browse/all/romabzcat.html}), each 5BZCAT source is classified in one of the four following categories based on its multiwavelength properties.

\begin{enumerate}
    \item 5BZB: BL Lac objects, used for AGNs with a featureless optical
        spectrum, or having only absorption lines of galaxian origin and weak
        and narrow emission lines.
    \item 5BZG: sources, usually reported as BL Lac objects in the literature,
        but having a SED with a significant
        dominance of the galaxian emission over the nuclear one. Before including a source in a follow-up analysis aimed at understanding how its SED could look like in the TeV band, we need to ensure that the SED emission is not significantly contaminated by the host or, in the X-ray band, by non--jet emission. While this should not be a problem for BZB sources, in BZGs the emission from the optical to the X-ray band could be significantly affected by non--jet processes, which in turn would make both the X-ray to radio flux ratio and the synchrotron peak measurements unreliable. In this regard, a specific class of sources stands out the most: clusters of galaxies, whose radio and X-ray properties can in many cases be similar to those of the BZG sources. For example, \citet{green17} reported a list of 41 BZGs whose properties (most importantly, the presence of extended emission in the X-rays and the fraction of polarization in the radio) indicated that the emission of these sources is dominated by the cluster. We thus exclude these 41 objects from our analysis.
    \item 5BZQ: Flat Spectrum Radio Quasars (FSRQs), with an optical spectrum showing
        broad emission lines and dominant blazar characteristics.
    \item 5BZU: blazars of uncertain type (BCUs), adopted for a small number of sources having peculiar characteristics (for example sources that have been observed transitioning from a FSRQ state to a BL Lac one, or vice versa) but nonetheless exhibiting blazar activity.
\end{enumerate}

We cross--matched the 5BZCAT with the \lat\ 14-year Source Catalog \citep[hereafter 4FGL-DR4,][]{abdollahi22,ballet23}, which contains 7194 $\gamma$-ray sources, 5788 of which are classified as point-like, that have been detected in 14 years of observations in the 50\,MeV--1\,TeV energy range with the \lat. To do so, we performed a positional cross-match between the 5BZCAT coordinates and the coordinates reported as ``counterpart coordinates'' in the 4FGL--DR4 catalog, since their accuracy is much higher than the one of the \lat\ sources themselves, coming from a radio, optical, or X-ray catalog. 
We find a \lat\ counterpart for 1772 5BZCAT sources, while the remaining 1748 do not have one. 
The median separation between the 5BZCAT sources and the 4FGL-DR4 counterparts is 0.11$^{\prime\prime}$, and 1725 out of 1772 objects (i.e., 97\,\% of the 4FGL--DR4 sources) have a positional separation smaller than 2$^{\prime\prime}$. 
We note that such a small separation is a direct consequence of the fact that the counterpart coordinates reported in the 4FGL-DR4 catalog are always the most accurate available; in the vast majority of cases, such coordinates coincide almost exactly with the 5BZCAT ones.

For the most part, the remaining sources with separation larger than 2$^{\prime\prime}$ have a 4FGL-DR4 counterpart positional uncertainty between 16 and 32$^{\prime\prime}$: these counterparts are generally RX sources, detected by the \textit{ROSAT} all-sky X-ray telescope \citep{voges99}, or NVSS sources, detected by the National Radio Astronomy Observatory (NRAO) Very Large Array (VLA) Sky Survey \citep{condon98}. Thus, the separation between the 5BZCAT source and the 4FGL--DR4 counterpart is well within the 4FGL--DR4 counterpart positional uncertainty.

\subsection{X-ray counterpart identification}\label{sec:x_ray}
As discussed in the introduction, the purpose of this work is to understand if among the blazars not detected by \lat\ can be found a population of sources that are TeV-emitters that could be detected by current or future Cherenkov telescopes. To test this possibility, we focus on the X-ray detected blazars, since X-ray emission can be an effective tracer of the TeV emission, as already discussed in the introduction of this work. In the following subsections, we briefly present the X-ray telescopes and catalogs that we use in this paper.

\subsubsection{\xmm\ and the 4XMM-DR13 catalog}\label{sec:xmm}
The \xmm\ telescope has been launched on December 10, 1999, and has been operational since February 2000. \xmm\ combines an excellent effective area in the 0.2--12\,keV band, a large ($\sim$30$^\prime$ diameter) field of view, and a good angular resolution ($\sim$5$^{\prime\prime}$ on--axis point-spread function, PSF). For this work, we use the most recent release of the \xmm\ Survey Science Catalogue, 4XMM-DR13\footnote{\url{http://xmmssc.irap.omp.eu/Catalogue/4XMM-DR13/4XMM_DR13.html}} \citep{traulsen20,webb20}. 
As reported in the 4XMM webpage, ``[t]he 4XMM-DR13 catalogue contains source detections drawn from 13243 \xmm\ EPIC observations, covering an energy interval from 0.2 keV to 12 keV. These observations were made between 2000 February 3 and 2022 December 31''. 4XMM-DR13 contains 656,997 unique sources detected over an area of $\sim$1328\,deg$^2$. The median 0.2--12\,keV flux of the 4XMM-DR13 population is $\sim$2.2 $\times$ 10$^{-14}$\,\flu; the median positional accuracy is $\sim$1.57$^{\prime\prime}$.

For each source, a wide variety of parameters is reported, among which we mention its coordinates; count rates; fluxes, and hardness ratios in different bands; number of times the source has been observed by \xmm; variability significance. 
The full list of parameters available for each source can be found in the dedicated webpage: \url{http://xmmssc.irap.omp.eu/Catalogue/4XMM-DR13/4XMM-DR13_Catalogue_User_Guide.html}. There are 313 5BZCAT sources (i.e., $\sim$9\,\% of the whole 5BZCAT sample) with a 4XMM-DR13 counterpart within 10$^{\prime\prime}$. 

\subsubsection{\cha\ and the 2CSC catalog}\label{sec:chandra}
The \cha\ telescope has been launched on July 23, 1999. \cha\ covers the 0.3--7\,keV energy range and is the only X-ray instrument with subarcsecond on-axis angular resolution (PSF$\sim$0.5$^{\prime\prime}$). The field of view of its ACIS-I array is 16.9$^\prime$ $\times$ 16.9$^\prime$.
The most recent release of the \cha\ Source Catalog, CSC version 2.0.1 (hereafter 2CSC) was released on November 24, 2020 \citep{evans20}. This catalog contains 317,167 unique sources detected over an area of $\sim$560\,deg$^2$: as for the \xmm\ catalog, the information on multiple parameters is reported for each source. The 2CSC can be accessed at \url{http://cda.cfa.harvard.edu/cscweb/index.do}.

There are 218 5BZCAT sources (i.e., $\sim$6\,\% of the whole 5BZCAT sample) with a 2CSC counterpart within 5$^{\prime\prime}$: with respect to \xmm\, we choose to use this more conservative maximum separation due to \cha\ much better angular resolution. For the purposes of this work, we rescaled the \cha\ 0.5--7\,keV fluxes to the 0.2--12\,keV band used in the \xmm\ catalog using an average value that takes into account the fact that FSRQs and low-- and intermediate--energy peaked BL Lacs have a typical X-ray photon index $\Gamma_{\rm X}\sim$1.6 and a flux correction factor $k_{\rm corr}$=$f_{\rm 0.2-12 keV}$/$f_{\rm 0.5-7 keV}$ = 1.5 \citep[see, e.g.,][]{langejahn20,middei22}, while high--energy peaked BL Lacs typically have a steeper photon index, $\Gamma_{\rm X}\sim$2.5 and $k_{\rm corr}$ = 1.7 \citep[see, e.g.,][]{middei22}. The \cha\ to \xmm\ flux correction factor we used is thus $k_{\rm corr}$=1.6. We report in Section~\ref{sec:x_ray_summary} the number of sources detected by both \cha\ and \xmm.

\subsubsection{\xrt\ and the 2SXPS catalog}\label{sec:xrt}
The Neil Gehrels \textit{Swift} Observatory was launched on November 20, 2004, and mounts three different telescopes to observe the sky in the optical and ultraviolet (Ultraviolet/Optical Telescope, or UVOT), in the hard X-rays (14--195\,keV; Burst Alert Telescope, or BAT), and in the soft X-rays (0.2--10\,keV; X-ray Telescope, or XRT). The XRT has an on-axis angular resolution of $\sim$18$^{\prime\prime}$ and a 23.6 $\times$ 23.6 arcmin$^2$ field of view.
For this work, we use the second \textit{Swift} X-ray Point Source (2SXPS) catalog of detections by \xrt\ used in Photon Counting (PC) mode in the 0.3-10 keV energy range. The 2SXPS catalog is based on the observations performed by \xrt\ between January 1st, 2005, and August 1st, 2018, covers an area of 3790 deg${2}$, and contains 206,335 unique sources. 

The 2SXPS catalog can be accessed from the webpage \url{https://heasarc.gsfc.nasa.gov/W3Browse/swift/swift2sxps.html}.  There are 1666 5BZCAT sources (i.e., $\sim$47\,\% of the whole 5BZCAT sample) with a 2SXPS counterpart within 10$^{\prime\prime}$: 1191 also have a 4FGL-DR4 counterpart, while 475 are X-ray only sources. We rescale the 0.3--10\,keV flux in the 2SXPS catalog to the 0.2--12\,keV band by using a correction factor  $k_{\rm corr}$=$f_{\rm 0.2-12 keV}$/$f_{\rm 0.3-10 keV}$ = 1.13, which is computed assuming a photon index $\Gamma_{\rm X}\sim$2, following an approach consistent with the one discussed for \cha.

\subsubsection{eROSITA and the eROSITA-DE Data Release 1 catalog}\label{sec:erosita}
eROSITA \citep[extended ROentgen Survey with an Imaging Telescope Array][]{predehl21} is an X-ray telescope mounted on the Spektrum
Roentgen Gamma (SRG) orbital observatory \citep{sunyaev21}, and covers the 0.2--8\,keV energy range, although with a significant decline in effective area at energies $>$2.3\,keV. SRG was launched on July 13, 2019. eROSITA is a whole sky survey instrument: in this work we however make use of the eRASS1 catalog \citep{merloni24} of the Western Hemisphere (which is, in Galactic coordinates, 359.9442$>$GAL$\_$LONG$>$179.9442), that uses data taken in the first six months of eROSITA observations, that have been completed in June 2020.

Since eROSITA is more sensitive at lower energies, while its effective area sharply declines at energies $\gtrsim$3\,keV, the eRASS1 band of reference for the flux computation
is the 0.2--2.3\,keV one. For the purposes of this work, the 0.2--2.3\,keV fluxes have therefore been rescaled to the 0.2--12\,keV ones by assuming a simple power law model with photon index $\Gamma$=2, which corresponds to a flux a correction factor $k \sim$ 1.76.

The 5BZCAT contains 1608 blazars\footnote{Four additional sources (5BZG J0439+0520, 5BZG J1108--0149, 5BZG J1119+0900, and 5BZG J1445+0039) are in the sample of misclassified clusters reported in \citet{green17} and are thus not included in our analysis.} in the 20,627\,deg$^2$ footprint of eRASS1. 
For consistency with the approach used in searching for counterparts in the other X-ray catalogs, we ran a cross-match analysis with 10$^{\prime\prime}$ maximum separation.We find that 1379 sources out of 1608 have a eROSITA counterpart (i.e., $\sim$39\,\% of the whole 5BZCAT sample).
The median offset between the 5BZCAT and eROSITA positions is $\sim$2.6$^{\prime\prime}$). As a reference, in \citet{gatuzz24}, which makes use of the not yet released eROSITA catalog of blazars, the maximum separation radius between the eROSITA position and the 5BZCAT one is 8$^{\prime\prime}$. 
 
When breaking down the sample of 1379 5BZCAT sources with eROSITA counterpart by class, we obtain the following distribution: 812 BZQs; 415 BZBs; 69 BZGs; 83 BZUs. As a reference, in \citet{gatuzz24} it is stated that the eRASS1 blazar catalog (Hammerich et al. in prep.) will include 843 FSRQ and 666 BL Lacs.

We then performed a further cross-match, this time focusing on the non-4FGL subsample: we matched the 653 eROSITA sources with the 572 sources with archival X-ray data. We break down our step-to-step approach to determine how many eRASS1 sources have a counterpart in one of the other X-ray archival catalogs used in this work

\begin{enumerate}
    \item First, we find that 247 out of 572 sources with archival X-ray data and without 4FGL counterpart lie in the Western half of the eROSITA sky and thus can in principle have a counterpart in the eRASS1 catalog).
    \item For the vast majority of these non--4FGL sources with available archival X-ray observations (218 out of 247, $\sim$88\,\%) we find a counterpart in the eRASS1 catalog.
    \item The 29 remaining sources that are in at least one archival catalog but not in the eROSITA one are mostly faint sources, with median 0.2-12 keV flux $f_{\rm 0.2-12keV}$=2.4 $\times$ 10$^{-13}$\,\flu; only seven sources have $f_{\rm 0.2-12keV}$$>$5 $\times$ 10$^{-13}$\,\flu. This provides us with a reasonable explanation of their non--detection in eRASS1.
    \item Finally, 435 objects are in the eROSITA catalog but not in the archival sample, and can thus be defined as ``eRASS1 only'' sources, as they are named in Table~\ref{tab:summary_X_ray_obs}. The vast majority of the non--4FGL, eRASS1 only blazars are BZQs (341); 52 are BZBs, 21 BZGs, and 21 BZUs.
\end{enumerate} 

\subsubsection{\NuSTAR\ and the Nublazar catalog}\label{sec:nustar}
The \NuSTAR\ telescope \citep{harrison13} was launched on June 13, 2012, and is the first telescope to focus photons at energies above 10 keV. \NuSTAR\ covers the 3 to 79 keV energy range, has a $\sim$13$^\prime$$\times$13$^\prime$ field of view, and a $\sim$58$^{\prime\prime}$ angular resolution (half power diameter). Given its limited field of view, \NuSTAR\ is mostly used as follow--up instrument of known sources, rather than as a survey instrument, although \NuSTAR\ surveys of several known fields have been taken across the years (e.g., COSMOS, \citealt{civano15}; the Extended \cha\ Deep Field South, \citealt{mullaney15}; the North Ecliptic Pole time-domain field, \citealt{zhao21,zhao24} )

During the years, \NuSTAR\ has targeted a significant number of blazars: for this work, we use as a reference the NuBlazar catalog \citep{middei22}, that contains 126 blazars observed by \NuSTAR\ since its launch to September 30, 2021. Several blazars in the catalog have been observed with \NuSTAR\ more than once, and the catalog contains the properties of 253 different \NuSTAR\ exposures. Out of these 126 blazars, 114 are reported in the 5BZCAT: all of them are also reported in at least one of the 0.3--10\,keV catalogs we presented in the previous paragraphs. Out of 114 sources, 88 have a 4FGL--DR4 counterpart, while 26 do not have one. In the rest of this work, we will always include the \NuSTAR\ information, when available, when analyzing the SEDs of our sources.

\subsection{Summary of the X-ray counterparts of the 5BZCAT sources}\label{sec:x_ray_summary}

We report in Table~\ref{tab:summary_X_ray_obs} the number of sources detected in each of the five X-ray catalogs, and in selected combinations of the five, as well as the fraction of objects which have or lack a counterpart in the \lat\ 4FGL-DR4 catalog. Overall, 2435 5BZCAT sources (i.e., $\sim$68\,\% of the whole 5BZCAT sample) have at least one X-ray counterpart: 1428 out of these 2435 also have a $\gamma$-ray counterpart in the 4FGL-DR4 catalog, while the remaining 1007 do not have one. We note that the excellent positional accuracy of the 5BZCAT sources ensured that every 5BZCAT source is uniquely associated to a single X-ray object, and there are no occurrences of multiple radio sources cross-matched to the same X-ray target. 
For this same reason, we expect the number of spurious associations to be small. In \citet{ajello20} it is reported that the expected false-positive counterpart rate for the 4FGL AGN population is 1.6\,\%, which means that just $\sim$ 23 objects out of 1428 might be wrongly associated. We expect this fraction to be a good proxy for the 1007 5BZCAT sources with no 4FGL counterpart as well. In fact, blazars are rare objects both in the radio and in the X-ray band, with number densities $\lesssim$1 even at the faintest X-ray fluxes and radio flux densities probed in this work \citep[see,e.g.,][and Figures 4 and 5 in \citealt{padovani07}]{padovani95}). This, combined with the excellent positional accuracy provided both by the radio and by the X-ray observations, as discussed in the previous sections, allows us to safely assume a false-positive counterpart rate $<$2\,\% (i.e., $<$20 sources) for the non--4FGL sample.

In the first part of this work we will focus on the subsample of 464 sources, among the 2435 5BZCAT objects with at least one X-ray counterpart, that are reported in either the \xmm\ or in the \cha\ source catalogs: specifically, 67 sources have been observed by both \cha\ and \xmm\; 246 only by \xmm; 151 only by \cha. Out of these 464 sources, 271 (58\,\%) have a 4FGL-DR4 counterpart, while 193 (42\,\%) do not have one. In Section~\ref{sec:4FGL_vs_no4FGL} will use this subsample of sources which are also, on average, the targets with the best multi-wavelength coverage, to get a first understanding of the parameter space covered by the non--4FGL\footnote{From now on, when we talk about 4FGL or non--4FGL samples we are implicitly referring to the 4FGL--DR4 catalog.} population with respect to the 4FGL one. Later in the paper, in Section~\ref{sec:no4FGL_multiwave} we will then include in our analysis also the 814 non--4FGL sources with an X-ray counterpart in either the 2SXPS or in the eRASS1 catalogs.

\subsection{The TeVCAT catalog}\label{sec:TeVCAT}
 The TeVCAT - TeV Gamma-Ray Source Catalog\footnote{\url{https://heasarc.gsfc.nasa.gov/w3browse/all/tevcat.html}} \citep{wakely18} is a constantly updated online catalog of known VHE objects, which is, of sources detected at energies E$\gtrsim$50\,GeV. As of May 7, 2024, 77 5BZCAT sources were reported in the TeVCAT: all of them are also reported in the 4FGL-DR4 catalog. 72 out of these 77 sources have a redshift measurement. In the rest of the paper, we will systematically compare the properties of the sources in our sample to those of the known TeV emitters.

\begin{table*}
\centering
\renewcommand*{\arraystretch}{1.2}
\begin{tabular}{ccc|ccc}
\hline
\hline
Catalog & X-ray Instrument & Area Covered & Sources & With \lat\ & Without \lat\ \\
        &                  & deg$^2$      &         &            &               \\
\hline
\hline
4XMM-DR13   & \xmm         & 1328  & 313   & 181 (58\,\%)  & 132 (42\,\%)  \\ 
2CSC        & \cha         & 560   & 218   & 131 (60\,\%)  & 87 (40\,\%)   \\
2SXPS       & \xrt         & 3790  & 1666  & 1191 (71\,\%) & 475 (29\,\%)  \\
eRASS1      & eROSITA      & 20627 & 1379  & 726 (52\,\%)  & 653 (48\,\%)  \\
NuBlazar    & \NuSTAR      & 6     & 114   & 88 (77\,\%)    & 26 (23\,\%)  \\
\hline
\multicolumn{3}{c|}{4XMM-DR13 or 2CSC}      &  464  & 271 (58\,\%)  & 193 (42\,\%)  \\
\multicolumn{3}{c|}{2SXPS, no 4XMM or 2CSC}             & 1347  & 968 (72\,\%)  & 379 (28\,\%)  \\
\multicolumn{3}{c|}{eRASS1 only}           & 624   & 189 (30\,\%)  & 435 (70\,\%)  \\
\hline
\multicolumn{3}{c|}{Overall}                & 2435 & 1428 (59\,\%)  & 1007 (41\,\%)  \\
\multicolumn{3}{c|}{Of which in TeVCAT}     & 77   & 77 (100\,\%)   & 0 (0\,\%)     \\
\hline
\hline
\end{tabular}\caption{Summary of 5BZCAT sources detected in different catalogs of X-ray sources. For each catalog, or combination of catalogs, we also report the number of sources with and without a counterpart in the \lat\ 4FGL-DR4 catalog of sources detected in the 50\,MeV--1\,TeV energy range.}\label{tab:summary_X_ray_obs}
\end{table*}

\begin{figure*}[ht]
\begin{minipage}{0.51\textwidth} 
 \centering 
 \includegraphics[width=1\textwidth]{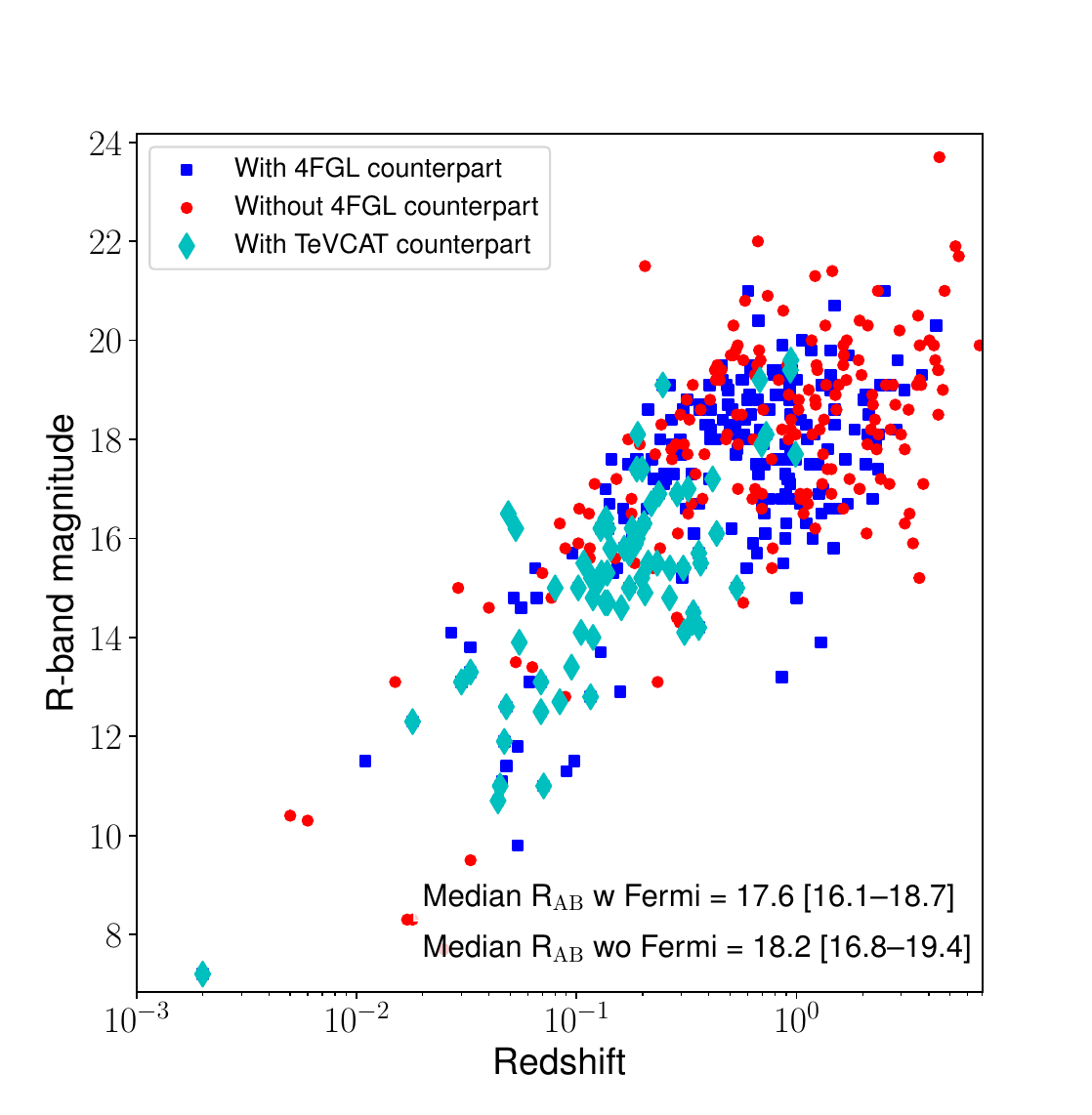} 
 \end{minipage} 
\begin{minipage}{0.48\textwidth} 
 \centering 
 \includegraphics[width=1\textwidth]{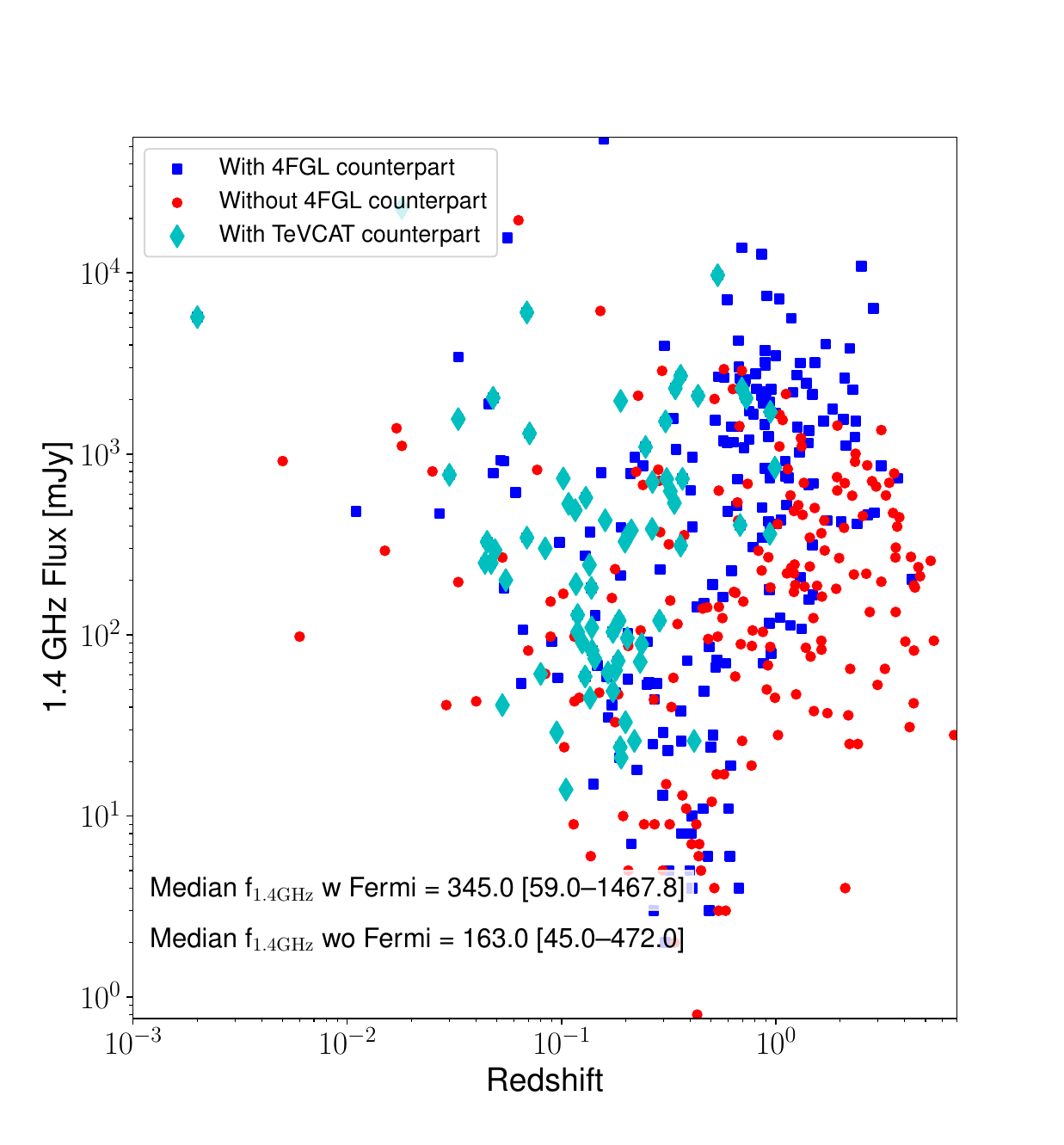}
 \end{minipage}
 \begin{minipage}{0.49\textwidth} 
 \centering 
 \includegraphics[width=1\textwidth]{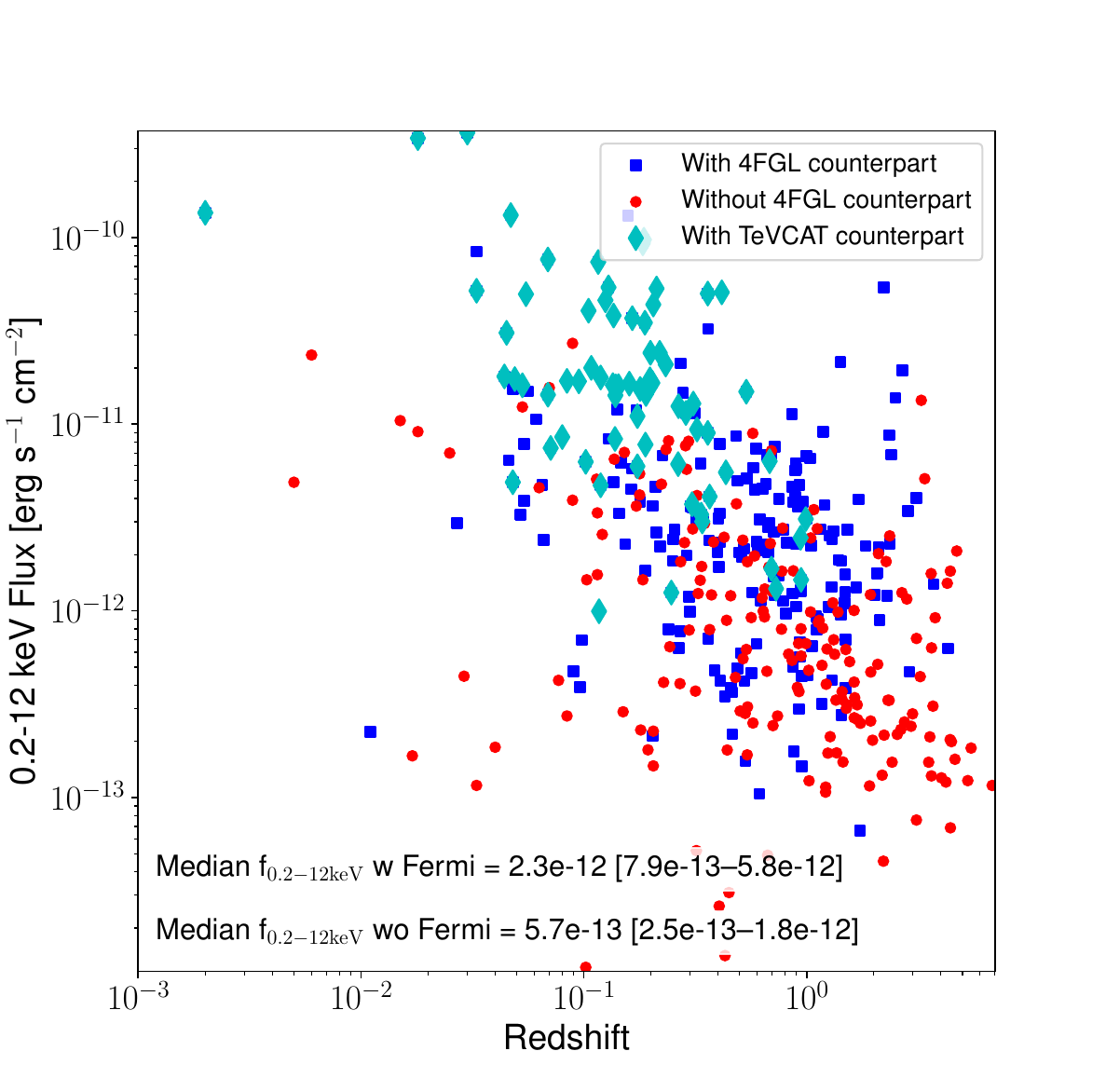} 
 \end{minipage} 
\begin{minipage}{0.49\textwidth} 
 \centering 
 \includegraphics[width=1\textwidth]{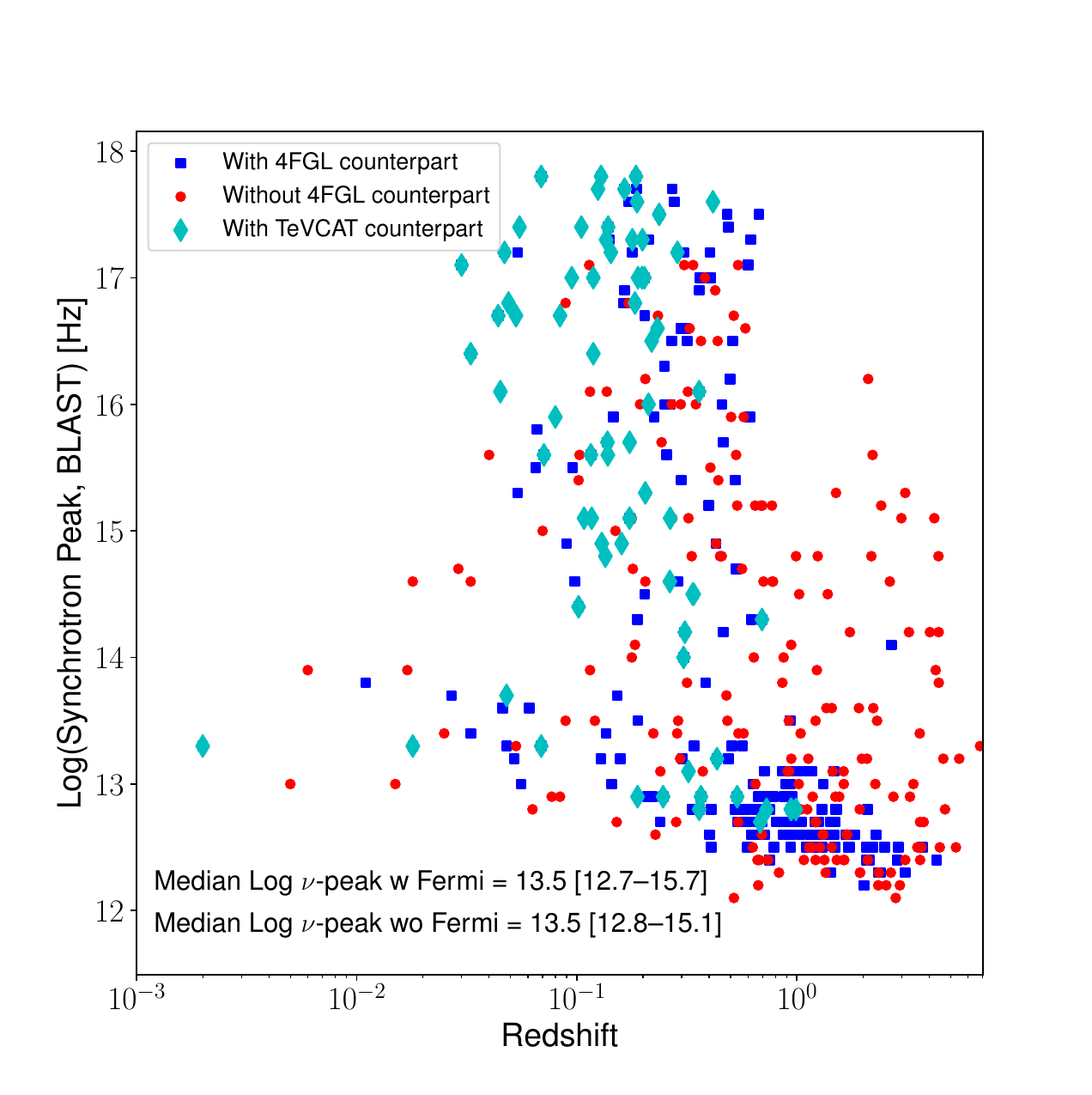}
 \end{minipage}
\caption{\normalsize 
R-band magnitude (top left panel), 1.4\,GHz flux density (top right panel), 0.2--12\,keV flux (bottom left panel) and BlaST--derived synchrotron peak frequency (bottom right panel) as a function of redshift for sources with a counterpart either in the 4XMM--DR13 catalog or in the 2CSC, with (blue squares) and without (red circles) a counterpart in the 4FGL-DR4 \lat\ catalog. As a reference, we also include in the plot the 72 sources with redshift information and a counterpart in the TeVCAT catalog (cyan diamonds), which are therefore known TeV--emitters. In the plot we also include the median values of each parameter for both the 4FGL and the non--4FGL sample: we include in square parentheses the first and third quartiles of each subsample.
}\label{fig:multiwave_Gamma_vs_no_Gamma}
\end{figure*}

\section{Multi-wavelength properties of the X-ray detected 5BZCAT sources}\label{sec:multiwave}
In this section, we initially study the multi-wavelength properties of the 464 sources that are reported in either the \xmm\ or in the \cha\ source catalogs, and in particular we compare the subsample of 271 sources having a \lat\ 4FGL counterpart with the one of 193 sources without one. The purpose of this first comparison is to understand if within the non--$\gamma$--ray population are included sources whose properties make them good candidates for a detection in the TeV band.

We report in Figure~\ref{fig:multiwave_Gamma_vs_no_Gamma} the distribution as a function of redshift of four parameters.

\begin{enumerate}
    \item R-band magnitude (top left panel; from the 5BZCAT).
    \item 1.4\,GHz flux density (top right panel; from the 5BZCAT).
    \item 0.2--12\,keV flux (bottom left panel; from the 4XMM and 2CSC catalogs).
    \item Synchrotron peak frequency (bottom right panel). This parameter has been computed using the VOU-BLazars tool \citep{chang20}, which allows one to quickly generate the SED of a given sources, and the BlaST \citep[Blazar Synchrotron Tool;][]{glauch22}, which is a deep neural network-based tool based on the work of \citet{lakshminarayanan16} which can be used to estimate the frequency of the synchrotron peak in a blazar SED.
\end{enumerate}

As it can be seen, a main trend can be noticed in the plots showing the brightness of the sources at different wavelengths: the non-4FGL detected sources tend to be slightly fainter than the 4FGL ones, although this difference in more prominent in certain regions of the electromagnetic spectrum than it is in others.
In the optical band (Figure~\ref{fig:multiwave_Gamma_vs_no_Gamma}, top left panel), for example, we observe only a mild difference between the two samples. The 4FGL--detected population has a median magnitude  R$_{\rm AB, 4FGL}$ = 17.6, with the first and third quartile of the population lying within R$_{\rm AB}$ = 16.1 and R$_{\rm AB}$ = 18.7, while the non--4FGL sample has R$_{\rm AB, no-4FGL}$ = 18.2 with the first and third quartile of the population lying within R$_{\rm AB}$ = 16.8 and R$_{\rm AB}$ = 19.4. To increase the paper readability, in the rest of the text we will report the first and third quartile values of a parameter between square brackets, as follows: R$_{\rm AB, no-4FGL}$ = 18.2 [16.8--19.4].

A more prominent discrepancy is instead observed in the radio part of the SEDs. In the 1.4\,GHz band (Figure~\ref{fig:multiwave_Gamma_vs_no_Gamma}, top right panel), the median flux density of 4FGL--detected sources is $f_{\rm 1.4 GHz,4FGL}$ = 345 [59 -- 1468] mJy. 
The median flux density of non--4FGL--detected sources is instead $f_{\rm 1.4 GHz,no-4FGL}$ = 163 [45 -- 472] mJy. Thus, the high--radio flux density end of the overall sample is dominated by the 4FGL--detected sources.

Finally, in the 0.2--12\,keV band, the median flux of 4FGL--detected sources is f$_{\rm 0.2-12 keV,4FGL}$ = 2 $\times$ 10$^{-12}$ [8 $\times$ 10$^{-13}$ --  6 $\times$ 10$^{-12}$] \flu.
For the non--4FGL population the median X-ray flux value is f$_{\rm 0.2-12 keV,no-4FGL}$ = 6 $\times$ 10$^{-13}$ [ 3 $\times$ 10$^{-13}$ -- 2 $\times$ 10$^{-12}$] \flu, so that the median flux value is a factor $\sim$4 smaller than the median X-ray flux of the 4FGL--detected subsample. 

The difference in brightness between the two samples is at least partly caused by the fact that the non 4FGL population includes a broader fraction of the high--z population. More in detail, the 4FGL sample median redshift is $z_{\rm 4FGL}$ = 0.53 [0.21 -- 0.94], while the non--4FGL sample median redshift is z$_{\rm no-4FGL}$ = 0.83 [0.32 -- 1.92]. Notably, 84 out 134 sources at $z>$1 (63\,\% of the subsample) do not have a 4FGL counterpart.

We note that, while the 4FGL sources are on average brighter in the X-ray and radio band than the non--4FGL ones, there is instead no significant evidence of a different trend in the 0.2--12\,keV to 1.4 \,GHz flux ratio as a function of redshift: the 4FGL sample has median ratio $r_{\rm X-to-Radio,4FGL}$=280 [101 -- 2507].
The non--4FGL sample, instead, has $r_{\rm X-to-Radio,no-4FGL}$=261 [84 -- 1124]. 
This ratio has been applied in the past to select potential candidate TeV emitters \citep[the so-called  extreme highly peaked BL Lac objects, or EHBLs; see, e.g.,][]{costamante01,bonnoli15}, since it can be a reliable indicator of a high synchrotron peak frequency \citep[and consequently of a inverse Compton peak in the TeV range; see, e.g.,][]{costamante18,foffano19,righi19}: we will further make use of this parameter in the following sections.

Finally, and in agreement with the point we just made on the use of the X-ray to radio flux ratio as a proxy of the SED shape, we do not observe a clear difference in the synchrotron peak distributions, with median Log($\nu_{\rm Peak}$)=13.5 in both populations, and fairly consistent first and third quartile ranges as well: Log($\nu_{\rm Peak, 4FGL}$) = [12.7--15.7], and Log($\nu_{\rm Peak, no-4FGL}$) = [12.8--15.1].

\subsection{4FGL vs non--4FGL multiwavelength properties by source class}\label{sec:4FGL_vs_no4FGL}
To further explore the properties of our sample, we search for different trends between the source classes reported in the 5BZCAT catalog and reported in Section~\ref{sec:sample}. As it is shown in Table~\ref{tab:properties_class}, our sample contains 227 FSRQs (or BZQs, in the 5BZCAT nomenklature); 143 BL Lacs (BZB); 43 BL Lacs where the SED emission shows a significant contribution from the host galaxy (BZG); and 51 blazars of uncertain type (BZU). A first thing to notice is that the 4FGL and non--4FGL samples are not homogeneous in terms of the fraction of sources belonging to a given class. More specifically, 44\,\% of the sources in the 4FGL subsample are classified as BZB, 39\,\% as BZQ, 10\,\% as BZU, and 8\,\% as BZG. The non--4FGL sample is instead dominated by the BZQ sources, that make 63\,\% of it, while the remaining 37\,\% is divided between BZBs (13\,\%), BZUs (12\,\%) and BZGs (11\,\%). This different composition of the two samples significantly contributes to their different redshift distribution, since (as shown in Table~\ref{tab:properties_class}) the higher redshift population is made almost entirely of BZQs.

As it is shown in Table~\ref{tab:properties_class}, the comparison between 4FGL--detected and --undetected sources shows for each of the four classes behaviors that are generally consistent with those of the overall population: the $\gamma$--ray population is on average brighter in both the radio and in the X-rays, while the trend is less significant in the optical band, particularly for the BZQ, BZG, and BZU population.

Finally, BZBs, BZGs and BZQs have similar distribution of the synchrotron peak frequency between 4FGL--detected and --undetected sources, while we observe a different trend in the BZU population. In fact, BZUs with a \lat\ detection have a median synchrotron peak frequency,  Log($\nu_{\rm Peak, BZU, 4FGL}$)=13.0, which is almost one order of magnitude lower than the median synchrotron peak frequency for BZUs without \lat\ counterpart (Log($\nu_{\rm Peak, BZU, 4FGL}$)=13.9). It is widely known that FSRQs generally peak at much lower frequencies than BL Lacs: this is clearly observable in the sample studied in this work, as well, since the median synchrotron peak frequency for the 227 FSRQs in our sample is Log($\nu_{\rm Peak, BZQ}$)=12.8, with standard deviation $\sigma_{\rm Log(\nu Peak, BZQ)}$=0.8, while for the 186 sources that make the BL Lac population (both BZB and BZG) we measure a median synchrotron peak frequency Log($\nu_{\rm Peak, BZB, BZG}$)=15.6, with standard deviation  $\sigma_{\rm Log(\nu Peak, BZB, BZG)}$=1.4.
Consequently, the discrepancy we measure in the BZU population may hint at the fact that the \lat--detected sample of BZUs contains a larger fraction of FSRQs with respect to the non--LAT--detected sample.

\begin{table*}
\centering
\scalebox{0.56}{
\renewcommand*{\arraystretch}{1.2}
\begin{tabular}{ccc|cccccc|cccccc}
\hline
\hline
      & & & \multicolumn{6}{c|}{With 4FGL-DR4 counterpart} & \multicolumn{6}{c}{Without 4FGL-DR4 counterpart} \\
Class & N$_{\rm src}$ & N$_{\rm TeV}$ & N$_{\rm src}$ & $z$ & R$_{\rm AB}$ & f$_{\rm 1.4 GHz}$ &  f$_{\rm 0.2-12 keV}$ & Log($\nu_{\rm Peak}$) & N$_{\rm src}$ & $z$ & R$_{\rm AB}$ & f$_{\rm 1.4 GHz}$ &  f$_{\rm 0.2-12 keV}$ & Log($\nu_{\rm Peak}$) \\ 
      &    &   &        &               &          & mJy               & \flu                  & Hz                   &         &      &          & mJy               & \flu                  & Hz             \\
\hline
\hline
\multirow{ 2}{*}{FSRQ (BZQ)}      & \multirow{ 2}{*}{227 (49\,\%)} & \multirow{ 2}{*}{5} & \multirow{ 2}{*}{105 (39\,\%)} & 0.93 & 18.1 & 1151 & 2.2$\times$10$^{-12}$ & 12.7 & \multirow{ 2}{*}{122 (63\,\%)}  & 1.36 & 18.3 & 232 & 5.3$\times$10$^{-13}$ & 13.1 \\
                                  &              &    &              & [0.62--1.48] & [16.9--19.0]  & [411--2197]    & [1.1--4.5]$\times$10$^{-12}$ & [12.5--12.9]  &   & [0.75--2.40]  & [17.0--19.4] & [96--618]    & [2.4--14.4]$\times$10$^{-13}$   & [12.5--14.0] \\
\multirow{ 2}{*}{BLL (BZB)}       & \multirow{ 2}{*}{143 (31\,\%)} & \multirow{ 2}{*}{27} & \multirow{ 2}{*}{118 (44\,\%)} & 0.30 & 17.6 &   62 & 2.1$\times$10$^{-12}$ & 15.6 &  \multirow{ 2}{*}{25 (13\,\%)}  & 0.38 & 19.4 & 12 & 3.2$\times$10$^{-13}$ & 15.9 \\
                                  &              &    &              & [0.17--0.45] & [16.7--19.2]  & [12--169]    & [0.3--5.1]$\times$10$^{-12}$ & [14.5--16.7]  &   &  [0.31--0.50]  & [18.4--19.8] & [5--17]    & [1.8--7.9]$\times$10$^{-13}$   & [15.4--16.5] \\
\multirow{ 2}{*}{BLL-host (BZG)}  &  \multirow{ 2}{*}{43 (9\,\%)} &  \multirow{ 2}{*}{3} &  \multirow{ 2}{*}{21 (8\,\%)}  & 0.14 & 16.1 &   92 & 2.4$\times$10$^{-12}$ & 15.8 &  \multirow{ 2}{*}{22 (11\,\%)}  & 0.12 & 16.0 & 45 & 4.4$\times$10$^{-13}$ & 15.5 \\
                                  &              &    &              & [0.07--0.20]  & [14.0--17.5]  & [53--324]    & [0.8--3.9]$\times$10$^{-12}$ & [14.6--17.2]  &   &  [0.08--0.20]  & [14.8--17.6] & [28--165]    & [2.3--24.2]$\times$10$^{-13}$   & [14.6--16.1] \\
\multirow{ 2}{*}{BCU (BZU)}       &  \multirow{ 2}{*}{51 (12\,\%)} &  \multirow{ 2}{*}{5} &  \multirow{ 2}{*}{27 (10\,\%)} & 0.68 & 17.5 &  629 & 4.1$\times$10$^{-12}$ & 13.0 &  \multirow{ 2}{*}{24 (12\,\%)}  & 0.44 & 17.1 & 156 & 2.0$\times$10$^{-12}$ & 13.9 \\
                                  &              &    &              & [0.05--0.87] & [14.0--18.9]  & [541--3467]    & [2.2--8.8]$\times$10$^{-12}$ & [12.7--13.5]  &   & [0.08--0.67]  & [13.5--19.2] & [85--829]    & [0.7--5.0]$\times$10$^{-12}$   & [13.0--14.9] \\
\hline
\multirow{ 2}{*}{Overall}         & \multirow{ 2}{*}{464}            & \multirow{ 2}{*}{40} & \multirow{ 2}{*}{271}      & 0.53 & 17.6         &  345       & 2.3$\times$10$^{-12}$        & 13.5 & \multirow{ 2}{*}{193}           & 0.83 & 18.2 & 163 & 5.7$\times$10$^{-13}$ & 13.5\\
                                  &                                  &                      &                            &  [0.21--0.94]    & [16.1--18.7] & [59--1468] & [0.8--5.8]$\times$10$^{-12}$ &   [12.7--15.7] & & [0.32--1.92] & [16.8--19.4]  & [45--472] & [2.5--18.0]$\times$10$^{-13}$ & [12.8--15.1]\\
\hline
\hline
\end{tabular}}
\caption{Median redshift (computed on the subsample of sources having a redshift measurement), R$_{\rm AB}$ magnitude, 1.4\,GHz flux density, 0.2--12\,keV flux, and logarithm of the synchrotron peak frequency for the four classes of sources reported in the 5BZCAT. The numbers in square parentheses are the first and third quartile values for each of the parameters. For each of the four classes, we further break down the sample in \lat--detected and --undetected sources. N$_{\rm src}$ is the number of sources in each of the four classes (in the parentheses we report the fraction of sources in the class with respect to the whole population), while N$_{\rm TeV}$ is the number of sources with a counterpart in the TeVCAT. We also report as a reference the median values for the whole sample of 464 sources with either \xmm\ or \cha\ counterpart.}\label{tab:properties_class}
\end{table*}

\begin{figure*} 
\begin{minipage}{0.46\textwidth} 
 \centering 
 \includegraphics[width=1\textwidth]{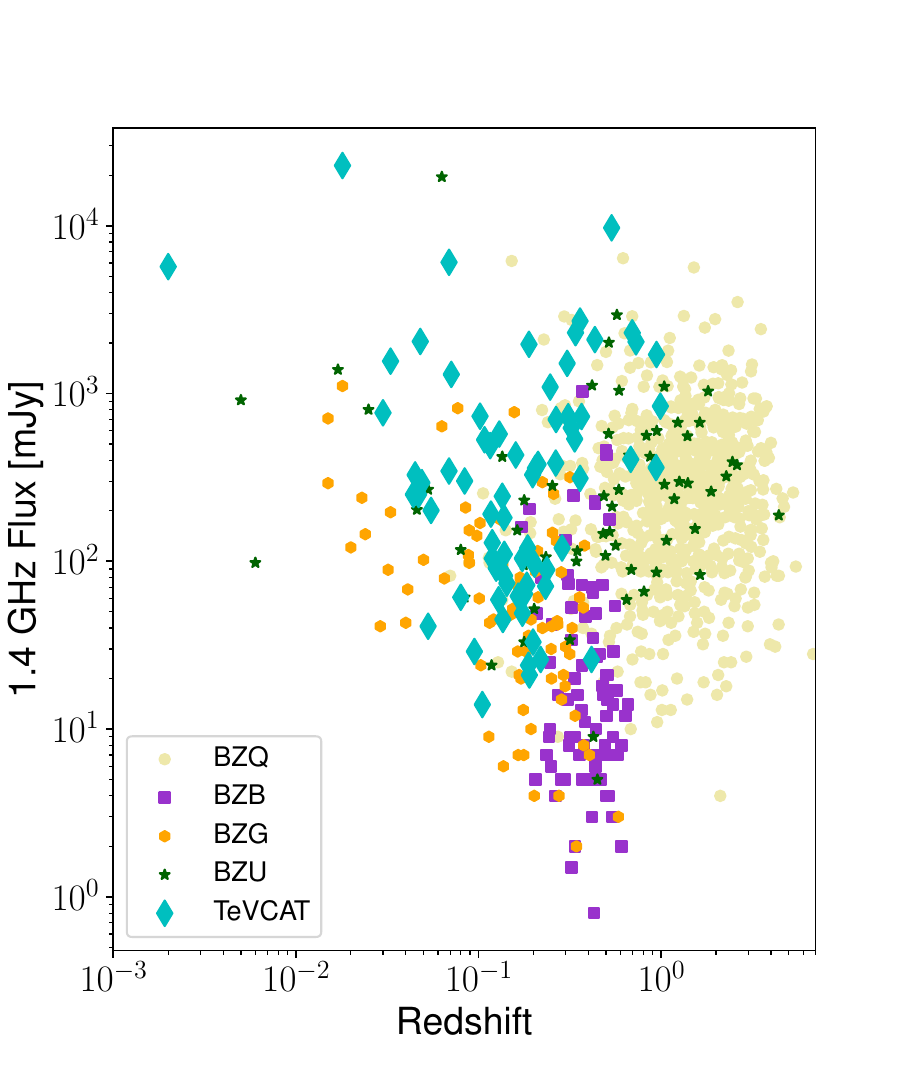}
 \end{minipage}
 \begin{minipage}{0.53\textwidth} 
 \centering 
 \includegraphics[width=1\textwidth]{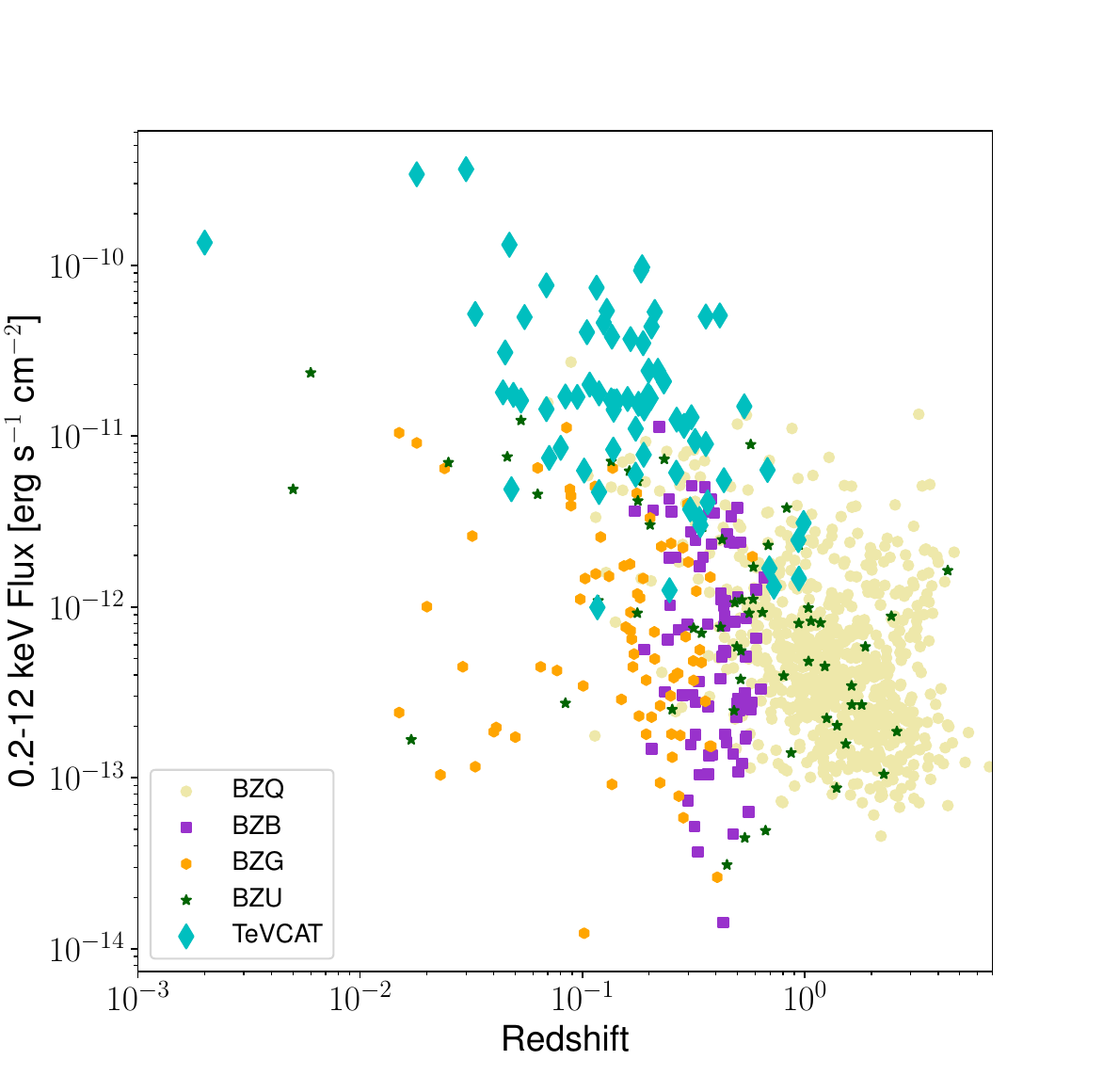} 
 \end{minipage} 
 \begin{minipage}{0.48\textwidth} 
 \centering 
 \includegraphics[width=1\textwidth]{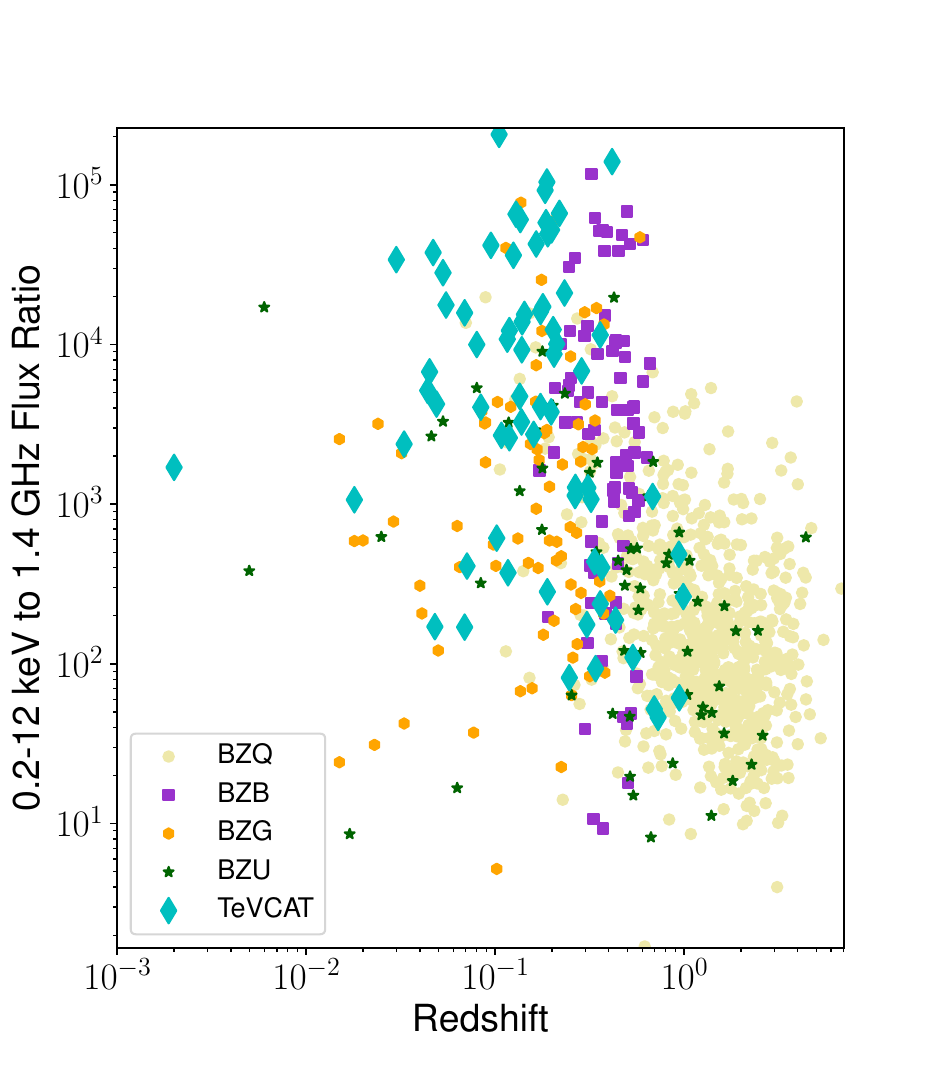} 
 \end{minipage} 
\begin{minipage}{0.51\textwidth} 
 \centering 
 \includegraphics[width=1\textwidth]{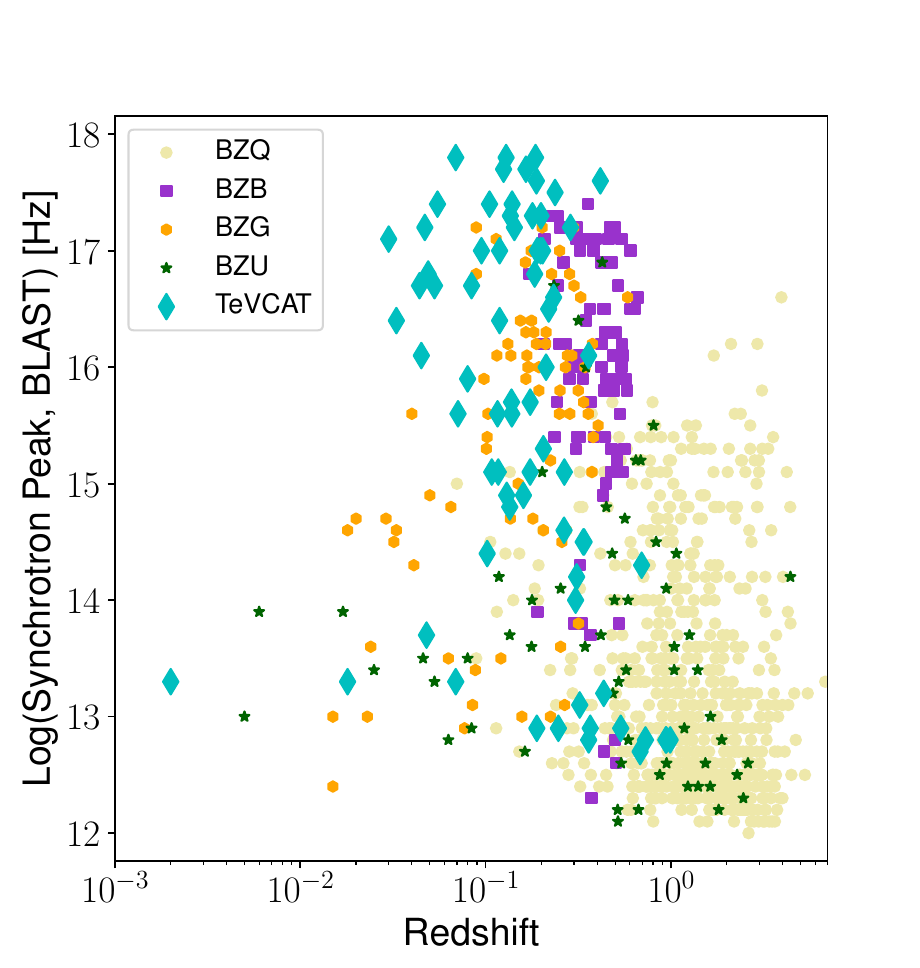}
 \end{minipage}
\caption{\normalsize 
R-band magnitude (top left panel), 1.4\,GHz flux density (top right panel), 0.2--12\,keV flux (bottom left panel) and BlaST--derived synchrotron peak frequency (bottom right panel) as a function of redshift for all the 593 5BZCAT sources without a counterpart in the 4FGL-DR4 \lat\ catalog and with at least one X-ray counterpart in the 4XMM--DR13, 2CSC, or 2SXPS catalogs. The different markers are linked to the 5BZCAT different blazar classes: BZQs (FSRQs) are plotted as pale gold circles, BZBs (BL Lacs) as violet squares, BZGs (BL Lacs with a significant contribution to the optical and infrared emission by the host galaxy) as orange hexagons, and BZU (blazars of uncertain classification) as green stars.
As a reference, we also include in the plot those sources with a counterpart in TeVCAT catalog (cyan diamonds) and are therefore known TeV--emitters.
}\label{fig:multiwave_no_Gamma_class}
\end{figure*}

\subsection{Properties of the known TeV emitters}\label{sec:TeVCAT_properties}
To further understand what are the typical properties of the extragalactic population emitting in the TeV band, we use the TeVCAT \citep{wakely18}. This, as mentioned earlier in the text, is a regularly updated catalog of known VHE objects, which is, of sources with a detection at energies $>$50\,GeV. In Figure~\ref{fig:multiwave_Gamma_vs_no_Gamma} we plot as cyan diamonds the 72 sources (out of 77) in our sample that have a TeVCAT counterpart and a redshift measurement: all of them are also detected by the \lat. As it is shown in the figure, the TeVCAT sources are located in the lower redshift, higher flux parameter space at all wavelengths. More in detail, in the X-ray band all but three TeVCAT sources have 0.2--12\,keV flux larger than 10$^{-12}$\,\flu. In the radio band, the TeVCAT sample covers a broader range of fluxes than in the X-rays; nonetheless, the median flux density of the TeVCAT population is $f_{\rm 1.4 GHz,TeVCAT}$ = 297\,mJy, is larger than the one of the 1242-sources sample of 4FGL--detected objects with X-ray counterpart ($f_{\rm 1.4 GHz,4FGL}$ = 222\,mJy). Finally, the median optical magnitude of the TeVCAT sample is R$_{\rm AB}$ = 15.4, over two orders of magnitude brighter than the one of the 4FGL--detected population (R$_{\rm AB, 4FGL}$ = 17.9). We also note that the median redshift of the TeVCAT sources is $z$=0.16\footnote{Four targets do not have a reliable redshift measurement.}, and only 9 sources have a redshift $z>$0.4 (six of these are classified as FSRQs, two as BZUs with low synchrotron peak frequency, and only one, TeV J0507+676, is a BL Lac). 

Finally, the synchrotron peak frequency distribution is shifted towards larger frequencies with respect to the overall population. More specifically, the TeVCAT subsample has median peak frequency Log($\nu_{\rm Peak, TeVCAT}$)=15.9. This is consistent with the fact that the vast majority of known TeV--emitting extragalactic sources are BL Lacs, rather than FSRQs.
Indeed, in our sample of 77 TeVCAT--detected sources, 61 are classified as BZB, 3 as BZG, and 5 as BZU, while only 8 are classified as BZQ. 
The numbers do not change significantly if we extend the comparison to the whole TeVCAT sample, which as of January 2024 contains $\sim$70 BL Lacs and only 10 BZQ sources, the latest of which -- OP 313 ($z$=0.997) -- has been detected by the Large-Sized Telescope (LST)--1 in December 2023 \citep{cortina23}.

\subsection{2SXPS and eRASS1 counterparts and the overall properties of the non--4FGL, X-ray detected population}\label{sec:no4FGL_multiwave}
To complete our analysis, and more specifically to obtain a census as complete as possible of those sources that do not have a \lat--detected counterpart but are instead detected in the X-rays, we include in our sample the 379 5BZCAT sources which lack a 4FGL-DR4 counterpart and have an X-ray counterpart in the \xrt\ 2SXPS catalog, as well as the 435 5BZCAT sources without a 4FGL counterpart and whose only X-ray counterpart is in the eROSITA eRASS1 catalog. The vast majority of these 814 sources (611 out of 814; 75\,\% of the 2SXPS or eRASS1--only subsample) are classified as BZQs, 106 (13\,\%) as BZB, 55 (7\,\%) as BZG, and 42 (5\,\%) as BZU. 

When combining these 814 sources with the 193 non--4FGL ones with either a 4XMM--DR13 or a 2CSC counterpart, we thus have a sample of 1007 5BZCAT sources without a 4FGL counterpart and with an X-ray detection. The sample has the following source class breakdown: 733 (73\,\%) BZQs, 131 (13\,\%) BZBs, 77 (8\,\%) BZGs, and 66 (7\,\%) BZUs.

We report in Figure~\ref{fig:multiwave_no_Gamma_class} the trend with redshift of the 1.4\,GHz flux density (top left), 0.2--12\,keV flux (top right), 0.2--12\,keV over 1.4\,GHz flux ratio (bottom left), and synchrotron peak frequency (bottom right) 
for all the 1007 5BZCAT sources without a counterpart in the 4FGL-DR4 \lat\ catalog and with at least one X-ray counterpart in the 4XMM--DR13, 2CSC, 2SXPS, or eRASS1 catalogs; each of these parameters have been derived in the same way presented in Section~\ref{sec:multiwave}. We also report a complete breakdown of the sample properties by source class in Table~\ref{tab:properties_class_no_4FGL}.

When analyzing the observed X-ray flux in the 0.2--12\,keV, we use the 4XMM--DR13 information when available (i.e., for 156 sources), the 2CSC one when the 4XMM--DR13 is not available (37 sources), the 2SXPS one when neither the 4XMM--DR13 nor the 2CSC one are available (379 sources), and the eRASS1 for the remaining 435 sources. 
As mentioned in Section~\ref{sec:x_ray}, the 2CSC, 2SXPS, and eRASS1 fluxes have been rescaled to the 4XMM-DR13 0.2--12\,keV band from the original 0.5--7\,keV, 0.3--10\,keV, and 0.2--2.3\,keV bands, respectively.

As it is shown in Figure~\ref{fig:multiwave_no_Gamma_class}, for all four parameters of interest, the parameter space populated by the known TeVCAT sources is also populated by a non negligible number of targets that belong to our non--4FGL sample. In particular, two classes of sources stand out significantly when analyzing the trends with redshift of the X-ray to radio flux ratio and with the synchrotron peak frequency: BZBs and BZGs, which is, BL Lacs sources where the broadband SED emission is dominated either by the jet emission (BZBs), or where the host galaxy emission significantly contributes to the overall SED (BZGs). 

Two parameters in particular can be used to parameterize this visual evidence: the  X-ray to radio flux ratio, and the synchrotron peak frequency. The median X-ray to radio flux ratio of the TeVCAT sample is $r_{\rm X-Radio}$ = 4485, with 55 out of 77 sources having $r_{\rm X-Radio}>$1000. The BZB sample has median $r_{\rm X-Radio}$ = 2111, with 54 out of 79 sources (68\,\%) having $r_{\rm X-Radio}>$1000, while the BZG sample has median $r_{\rm X-Radio}$ = 1084, with 28 out of 56 sources (50\,\%) having $r_{\rm X-Radio}>$1000. The TeVCAT, BZB, and BZG samples also have very similar median synchrotron peak frequencies: Log($\nu_{\rm Peak,TeV,Hz}$) = 15.9 [14.3 -- 17.0]; Log($\nu_{\rm Peak,BZB,Hz}$) = 16.0  [14.3 -- 16.5]; Log($\nu_{\rm Peak,BZG,Hz}$) = 15.6 [14.6 -- 16.2].

\begin{table*}
\centering
\scalebox{0.8}{
\renewcommand*{\arraystretch}{1.2}
\begin{tabular}{ccccccccc}
\hline
\hline
Class & N$_{\rm src}$ & $z$ & R$_{AB}$ & f$_{\rm 1.4 GHz}$ &  f$_{\rm 0.2-12 keV}$ & $r_{\rm X-Radio}$ & Log($\nu_{\rm Peak}$)\\ 
      &               &                                   &          & mJy          & $\times$10$^{-13}$ \flu                       &                   & Hz                   \\
\hline
\hline
FSRQ (BZQ)      & 733 (73\,\%) & 1.35 [0.88--2.10] & 18.5 [17.4--19.4] & 261 [119--469] & 4.3 [2.3--9.2]      & 125 [62--341]        & 12.9 [12.5--13.7] \\
BLL (BZB)       & 131 (13\,\%) & 0.42 [0.32--0.50] & 19.2 [18.6--19.6] &  19 [7--95]    & 3.8 [2.1--12.1]     & 1760 [213--7643]  & 15.8 [14.3--16.5] \\
BLL-host (BZG)  &   77 (8\,\%) & 0.18 [0.10--0.27] & 17.1 [14.9--18.1] &  53 [24--124]  & 5.6 [2.4--18.3]     & 730 [267--3184]  & 15.6 [14.6--16.2] \\
BCU (BZU)       &  66  (7\,\%) & 0.53 [0.20--1.05] & 18.1 [16.6--19.4] & 264 [107--560] & 8.0 [2.8--24.3]     & 296 [50--1015]       & 13.4 [12.7--14.2] \\
\hline
Overall         & 1007         & 1.12 [0.52--1.88] & 18.5 [17.4--19.3] & 204 [79--421]  & 4.5 [2.3--11.0]     & 170 [66--592]        & 13.2 [12.5--14.7] \\
\hline
\hline 
TeV population  &  77          & 0.17 [0.10--0.27] & 15.4 [14.3--16.3] & 294 [82--729]  & 160.6 [66.1--366.5] & 4737 [549--19327] & 15.8 [14.3--17.0] \\
\hline
\hline
\end{tabular}}
\caption{Median value with, in square parentheses, first and third quartile of the distribution for R$_{\rm AB}$ magnitude, 1.4\,GHz flux, 0.2--12\,keV flux, 0.2--12\,keV over 1.4\,GHz flux ratio, and logarithm of the synchrotron peak frequency for the four classes of sources reported in the 5BZCAT, for the 1007 X-ray detected sources without a 4FGL counterpart. The values are computed only for the sources for which a parameter measurement is available.
N$_{\rm src}$ is the number of sources in each of the four classes. We also report as a reference the median values for the whole sample of 1007 sources, as well as for the 77 TeVCat sources with an X-ray counterpart.}\label{tab:properties_class_no_4FGL}
\end{table*}

\begin{figure*} 
\begin{minipage}{0.5\textwidth} 
 \centering 
 \includegraphics[width=1\textwidth]{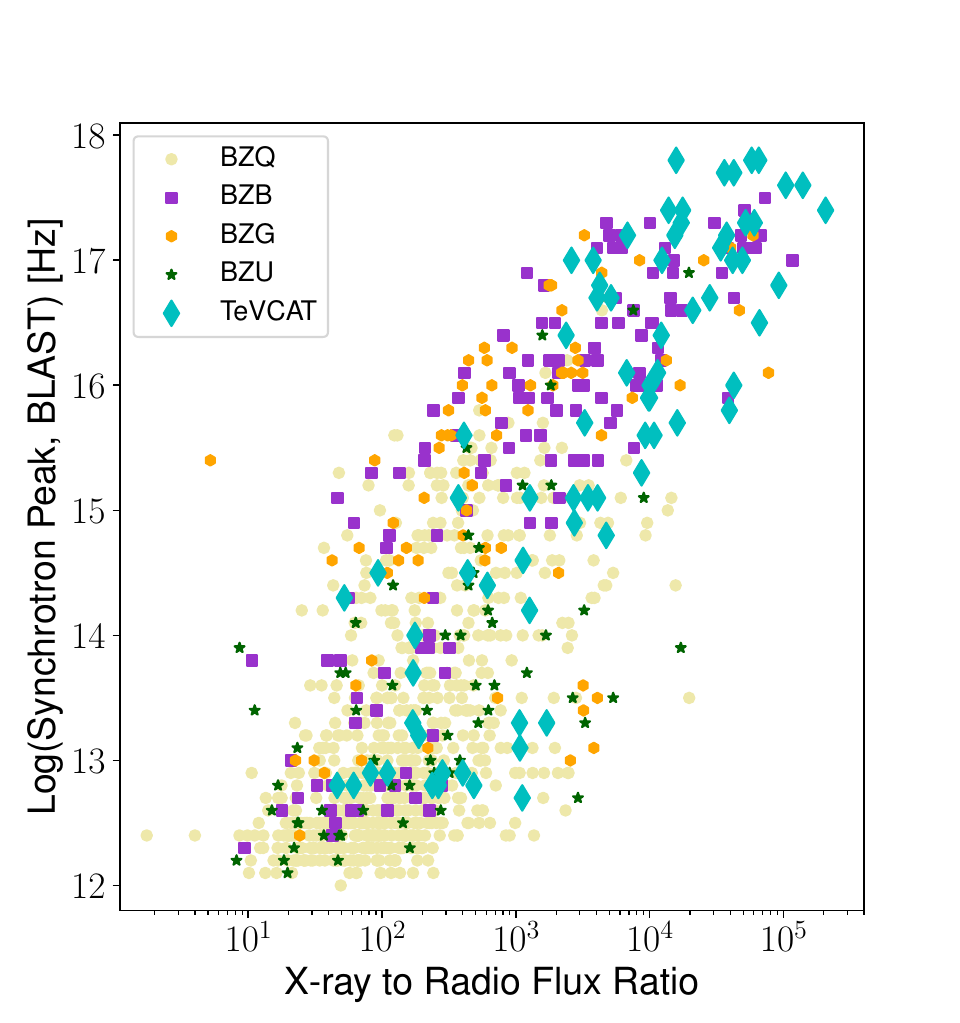} 
\end{minipage}
\begin{minipage}{0.49\textwidth} 
 \centering 
 \includegraphics[width=1\textwidth]{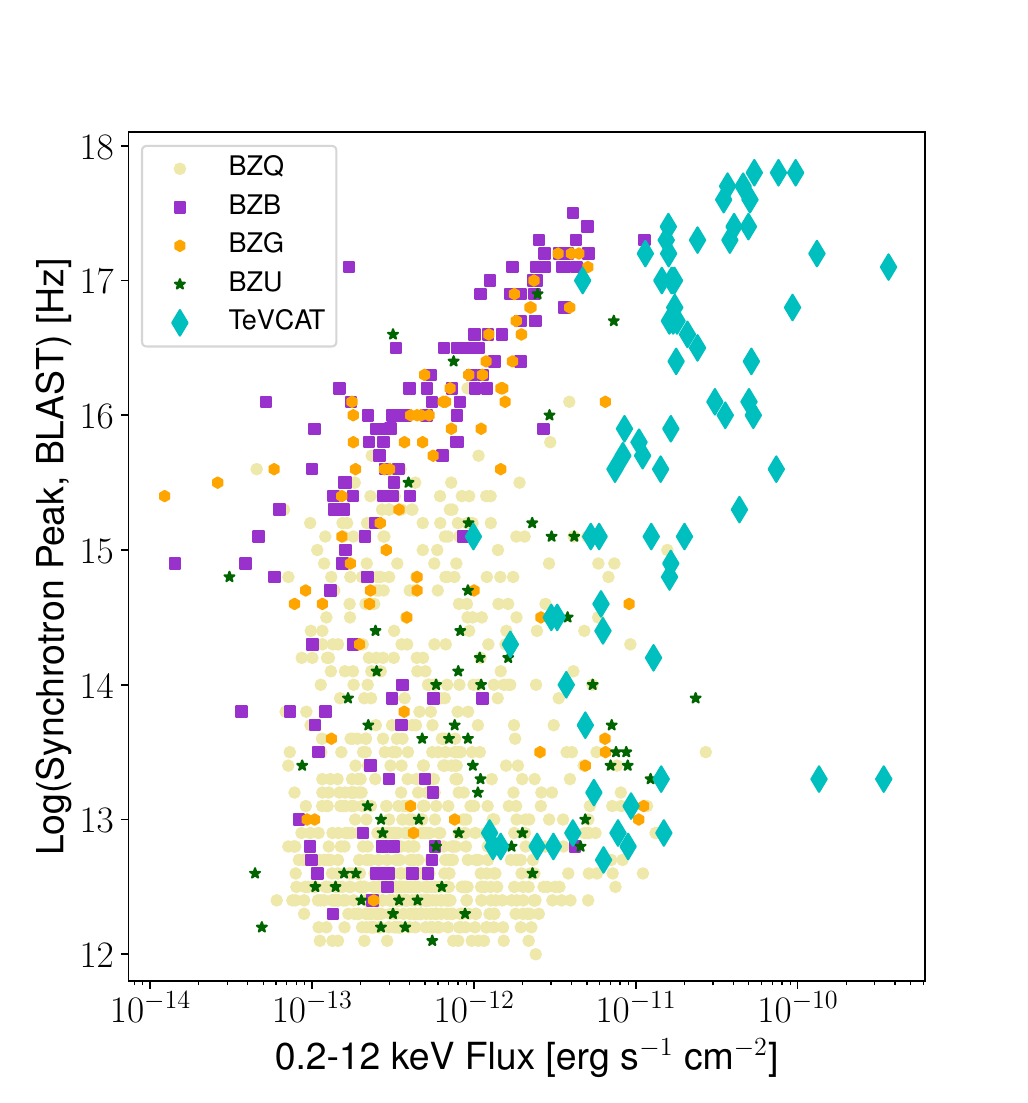} 
\end{minipage}
\caption{\normalsize 
BlaST--derived synchrotron peak frequency as a function of the ratio between the 0.2--12\,keV flux and the 1.4\,GHz one (left) and of the 0.2--12\,keV flux (right) for the 1007 5BZCAT sources without \lat\ counterpart and with an X-ray counterpart. The sources are color-coded by class, using the same coding described in Figure~\ref{fig:multiwave_no_Gamma_class}. 
}\label{fig:flux_ratio_vs_peak}
\end{figure*}

To further explore the overlap between the parameter space populated by known extragalactic TeV emitters and the X-ray--detected, non--4FGL 5BZCAT sample, we show in Figure~\ref{fig:flux_ratio_vs_peak}, left panel, the trend with the 0.2--12\,keV over 1.4\,GHz flux ratio of the SED synchrotron peak frequency. This plot shows even more clearly how a significant population of BZBs and BZGs lies in the same parameter space which contains the majority of the TeVCAT population, which is, at higher flux ratios and higher synchrotron peak frequencies. Specifically, 46 out of 77 TeVCAT sources (i.e., 61\,\% of the known TeV objects) have 0.2--12\,keV to 1.4\,GHz flux ratio $r_{\rm X-Radio}>$ 2 $\times$ 10$^{3}$ and synchrotron peak frequency Log($\nu_{\rm Peak}$)$>$15. 43 out of these 46 sources are classified as BZBs, and 3 as BZGs. In this same parameter space we have 70 5BZCAT sources that lack a 4FGL-DR4 counterpart: 41 are classified as BZB; 17 as BZG; 8 as BZQ; and 4 as BZU. 

We also report in Figure~\ref{fig:flux_ratio_vs_peak}, right panel, the trend with the 0.2--12\,keV flux of the SED synchrotron peak frequency. This plot immediately clarifies why the sources we selected lack (as of today) a TeV counterpart, despite being in the same region of the X-ray--to--radio flux ratio versus synchrotron peak parameter space of the majority of the known TeV emitters. As it can be seen (and as it was shown in Figure~\ref{fig:multiwave_no_Gamma_class}, top right panel), the X-ray flux of our targets is systematically fainter than the one of the known TeVCAT sources: more in detail, the median 0.2--12\,keV flux of the TeVCAT population is 1.6 $\times$ 10$^{-11}$ [6.6 $\times$ 10$^{-12}$ \flu\ -- 3.7 $\times$ 10$^{-11}$] \flu, while the median 0.2--12\,keV flux of the non--4FGL population is 4.5 $\times$ 10$^{-13}$ [2.3 $\times$ 10$^{-13}$ -- 1.1 $\times$ 10$^{-12}$] \flu. We report in Table~\ref{tab:properties_class_no_4FGL} a breakdown by source class, which shows that the median values do not change too significantly among the four source classes reported in the BZCAT, with the median X-ray flux varying between 3.8 $\times$ 10$^{-13}$ \flu\ for the BZB sample and 8.0 $\times$ 10$^{-13}$ \flu\ for the BZU sample.

Finally, we note that \citet{bonnoli15} also used a selection criterion on the flux ratio to select potential VHE blazars, which is, $r_{\rm 0.1-2.4keV/1.4GHz}>$ 10$^{4}$. The X-ray flux they used to compute the ratio was the ROSAT, 0.1--2.4\,keV one \citep{plotkin10}; when rescaled to the 0.2--12\,keV band flux used in this work assuming a photon index $\Gamma$=2 (as explained in Section~\ref{sec:x_ray}), the threshold corresponds to $r_{\rm 0.2-12keV/1.4GHz}\gtrsim$ 1.7 $\times$ 10$^{4}$. This criterion is met by only 21 out of 77 TeVCAT sources (i.e., $\sim$27\,\% of the sample), but is nonetheless effective in selecting a parameters space that is populated almost entirely by VHE emitters, particularly when including a secondary selection criterion based on the X-ray flux.

\section{Estimating the TeV emission of the population of X-ray bright, non--Gamma--ray detected targets}\label{sec:TeV_predictions}

\begin{figure*} 
 \centering 
 \includegraphics[width=0.8\textwidth]{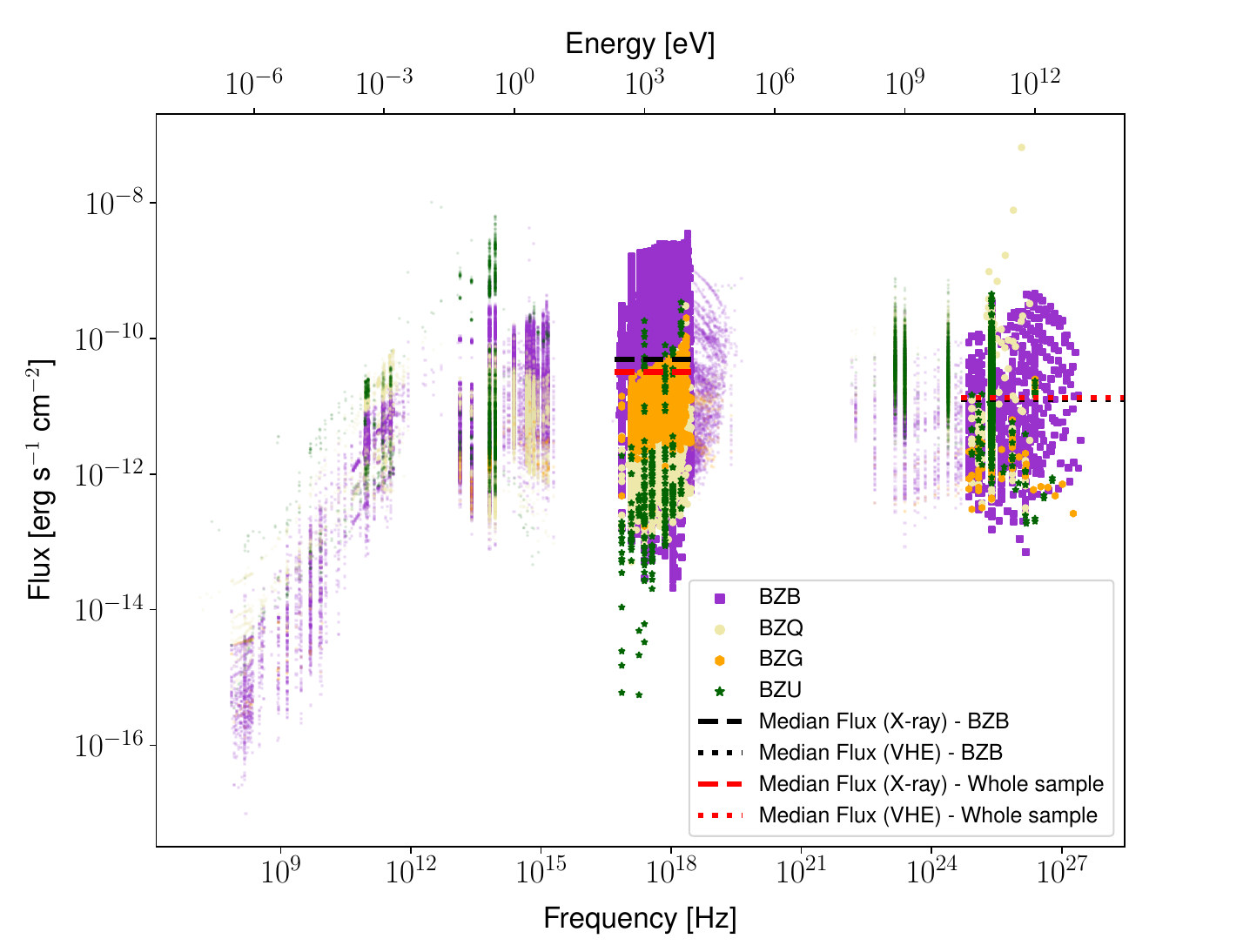} 
\caption{\normalsize 
Spectral energy distributions of the 77 TeVCAT sources analyzed in this work. The SEDs are color-coded by class, using the same coding described in Figure~\ref{fig:multiwave_no_Gamma_class}. We also show with black (red) dashed and dotted lines the median 0.2--12\,keV and 20\,GeV -- 300\,TeV fluxes of the BZB subsample (overall population).
}\label{fig:SED_TeV}
\end{figure*}

As shown in Figure~\ref{fig:flux_ratio_vs_peak}, right panel, and discussed in the previous section, our analysis led to the discovery of a population of sources that have multi-wavelength properties consistent with those of the known TeV blazars reported in the TeVCAT, but are generally fainter in the X-ray band. To get a first estimate of how bright in the TeV band could be, we use as a reference the 77 TeVCAT sources in our sample. In Figure~\ref{fig:SED_TeV} we report the SEDs of these objects, color-coded by source class. As it can be seen, the sources tend to share a similar SED shape, with the majority of the sources (with the exception of the BZQs) having a synchrotron peak frequency at energies larger than 10$^{15}$\,Hz, as already discussed in the previous section.

When analyzing the SEDs of known VHE emitters, we are interested the most in measuring fluxes in the X-rays (0.2--12\,keV) and in the VHE ($>$20\,GeV) bands, to determine if and how the two quantities are correlated. As mentioned earlier in the text, the vast majority of known TeV emitters are BL Lacs (BZB in the 5BZCAT classification): more in detail, our sample contains 61 BZBs, with a median 0.2--12\,keV flux value $f_{\rm X}$ = 4.9 $\times$ 10$^{-11}$\,\flu, and median VHE (20\,GeV -- 300\,TeV) flux value $f_{\rm VHE}$ 1.3 $\times$ 10$^{-11}$\,\flu. These two values are reported in Figure~\ref{fig:SED_TeV} as black dashed and dotted lines, respectively. In the same figure, we also report (as red dashed and dotted lines, respectively) the median 0.2--12\,keV and 20\,GeV -- 300\,TeV for the whole 77 TeVCAT sources, which are $f_{\rm X}$ = 3.2 $\times$ 10$^{-11}$\,\flu, and $f_{\rm VHE}$ = 1.3 $\times$ 10$^{-11}$\,\flu.

\begin{figure} 
 \centering 
 \includegraphics[width=0.49\textwidth,trim={1cm 1cm 0cm 0},clip]{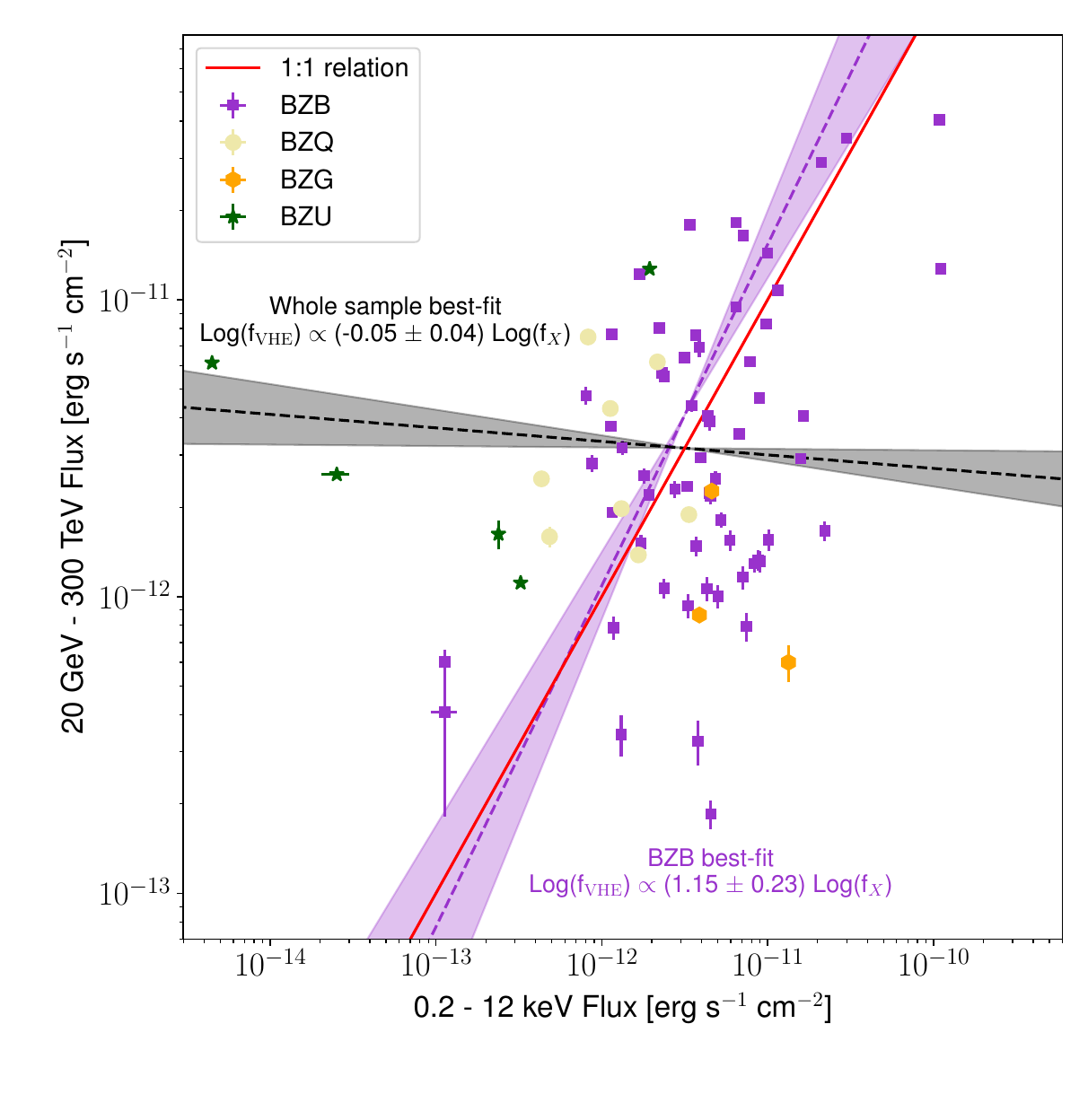} 
\caption{\normalsize 
Average 20\,GeV -- 300\,TeV flux as a function of the average 0.2--12\,keV one for 77 TeVCAT sources in our sample.
The data points are color-coded by class, using the same coding described in Figure~\ref{fig:multiwave_no_Gamma_class}. 
In the plot, we also include the best-fit linear regression between the logarithm of the VHE flux and the logarithm of the X-ray one, for the whole sample (dashed black line) and for the BZB subsample (violet dashed line). For reference, we also plot the 1:1 relation (solid black line).
}\label{fig:X_vs_TeV}
\end{figure}

To further explore the X-ray to VHE correlation, we plot in Figure~\ref{fig:X_vs_TeV} the average 20\,GeV -- 300\,TeV flux as a function of the average 0.2--12\,keV one for the 77 TeVCAT sources in our sample, once again color-coding them by source class. A few trends can be easily noticed: first of all, BZQs generally have VHE fluxes consistent with or, more often, larger than the X-ray ones. When performing a linear regression (in the Log-Log space) of the whole sample, we do not find a significant correlation between the two fluxes, with Log(f$_{\rm VHE}$) $\propto$ (-0.05 $\pm$ 0.04) Log(f$_{\rm X}$). A more prominent trend is instead observed when focusing on the BL Lac (BZB) population: in this case, while the distribution has a fairly large scatter, we measure a correlation Log(f$_{\rm VHE}$) $\propto$ (1.15 $\pm$ 0.23) Log(f$_{\rm X}$), consistent within the uncertainties with the 1:1 trend.

This means that, at the zero order, the X-ray flux can be treated as a fairly reasonable proxy of the VHE flux for BZB sources, and thus the plot shown in Figure~\ref{fig:flux_ratio_vs_peak}, right panel, can be treated as a synchrotron peak versus VHE Flux plot for the BZB subsample. It can then be seen that a non-negligible population of sources exist that might have VHE fluxes $\sim$10$^{-12}$\,\flu: these objects represent an intriguing pool of targets for dedicated observing campaigns with current facilities, such as MAGIC, HESS, VERITAS, or LST-1, and should be easily detected by future facilities such as the Cherenkov Telescope Array Observatory, whose sensitivity will improve the one of current facilities by almost one order of magnitude \citep{hofmann23}.

\section{Conclusions and future developments}\label{sec:conclusions}

We have analysed a sample of 2435 blazars reported in the 5BZCAT and having an X-ray counterpart in at least one of the \xmm, \cha, \xrt, or eROSITA point-like source catalogs, with the goal of determining if X-ray emission can be used as an effective proxy to find and characterize candidate TeV-emitting blazars that would be missed by simply extrapolating the source $\gamma$-ray emission. These are the main results of this work.

\begin{enumerate}
    \item We selected a sample of 464 sources with high--quality X-ray data from either \xmm\ or \cha\ observations to determine the multi-wavelength properties of the sources with a $\gamma$--ray counterpart in the 4FGL-DR4 catalog. We find that the non--4FGL sources are on average fainter both in the X-rays and in the radio with respect to the 4FGL--detected ones, but the two samples have similar X-ray--to--radio flux ratios, as well as synchrotron peak frequencies. Furthermore, as shown in Figure~\ref{fig:multiwave_Gamma_vs_no_Gamma}, when plotting these two parameters versus redshift, a significant fraction of the non--4FGL sources lies in a similar parameter space than the known TeV emitters.
    \item We then focused on the 1007 non--$\gamma$--ray detected population, to determine if there is a sample of X-ray sources that could be TeV emitters: for this purpose, we included in our sample also those sources detected only by \xrt, or eROSITA. We find that a large number of sources, mostly BL Lacs or BL Lacs with host-galaxy contribution to the SED, have large synchrotron peak frequency (Log$\nu_{\rm Peak/Hz}>$15) and X-ray to radio flux ratio ($r_{\rm X-Radio}>$1000), as reported both in Figure~\ref{fig:multiwave_no_Gamma_class} and Table~\ref{tab:properties_class_no_4FGL}. These
    two properties also characterize the vast majority of known TeV emitters: however, with respect to these known TeV emitters, our targets have X-ray fluxes $\sim$1 order of magnitude fainter.
    \item Finally, we computed the 0.2--12\,keV and 20\,GeV -- 300\,TeV fluxes for the known 5BZCAT TeV emitters, and determined the existence of a direct correlation between X-ray and TeV fluxes in the BL Lacs population. This trend can thus be used to estimate the VHE flux of our targets, to select sources for follow--up observations with current or future, more sensitive, Cherenkov telescopes, first and foremost the Cherenkov Telescope Array Observatory \citep{hofmann23}.
\end{enumerate}

The sample we selected can be used for a variety of purposes. Our group is currently working on the analysis of subclasses of targets with newly developed tools that allow one to fit the SED of a blazar with physically meaningful emission models (A. Iuliano et al. in preparation). We plan to test the SED fitting tool released by the Markarian Multiwavelength Data Center \citep[][\url{https://mmdc.am/}]{sahakyan24}, which is based on a convolutional neural network (CCN) whose parameters and technical implementation are presented in \citet{begue23}. Additionally, with this paper, we also make available the catalog of 1007 5BZCAT sources without 4FGL counterpart and with at least one detection in the 4XMM-DR13, 2CSC, 2SXPS, or eRASS1 catalogs: we present the properties of this catalog in the Appendix.

\section*{Acknowledgments}
The research activities described in this paper were carried out with contribution of the Next Generation EU funds within the National Recovery and Resilience Plan (PNRR), Mission 4 - Education and Research, Component 2 - From Research to Business (M4C2), Investment Line 3.1 - Strengthening and creation of Research Infrastructures, Project IR0000012 – ``CTA+ - Cherenkov Telescope Array Plus''.
This work makes use of Matplotlib \citep{hunter07} and NumPy \citep{harris20}.
S.M. thanks Lea Marcotulli for the helpful suggestions on blazar source catalogs, and Marco Ajello and Mauro Dadina for the useful discussions.

\bibliographystyle{aa}
\bibliography{Xray_blazars_biblio}

\appendix
\section{Catalog of the 1007 sources without a 4FGL counterpart}
Together with the paper, we make available the catalog of 1007 5BZCAT sources without 4FGL counterpart and with at least an X-ray counterpart. These are the columns of the catalog.

\begin{itemize}
    \item Column 1:  5BZCAT source ID, from \citet{massaro15}.
    \item Column 2--3: Right Ascension (R.A.) and Declination (Dec) of the 5BZCAT source.
    \item Column 4: Source redshift, when available. From the 5BZCAT.
    \item Column 5: Object classification from the 5BZCAT (BZB, BZG, BZQ, or BZU): see Section~\ref{sec:multiwave} for a detailed description of each of these classifications.
    \item Column 6: Optical, R-band magnitude of the source, from the 5BZCAT. This is, as reported in the 5BZCAT catalog description, the ``R-band magnitude of the blazar taken from the USNO-B1 Catalog, the r-band magnitude from SDSS DR10, or the magnitude in other bandpasses when these data are not available''.
    \item Column 7: 1.4\,GHz flux density of the source, from the 5BZCAT. This is, as reported in the 5BZCAT catalog description, the ``radio flux density of the blazar at 1.4\,GHz (from NVSS or FIRST), in mJy. Alternatively, if the 1.4\,GHz flux density is not known, the flux density at 0.843\,GHz from SUMSS is given''.
    \item Columns 8--10: source ID, R.A., and Dec from the \xmm\ 4XMM-DR13 X-ray catalog.
    \item Columns 11--13: source ID, R.A., and Dec from the \cha\  2CSC X-ray catalog.
    \item Columns 14--16: source ID, R.A., and Dec from the \xrt\  2SXPS X-ray catalog.
    \item Columns 17--19: source ID, R.A., and Dec from the eROSITA  eRASS1 X-ray catalog.
    \item Columns 20--22: 0.2--12\,keV observed X-ray flux and corresponding 90\,\% confidence level lower and upper uncertainties, from the first catalog from which the source is detected, in this order: 4XMM-DR13, 2CSC, 2SXPS, eRASS1. For \cha, \xrt, and eROSITA the fluxes have been rescaled to the 0.2--12\,keV band following the approach presented in Section~\ref{sec:x_ray}.
    \item Column 23: 0.2--12\,keV to 1.4 GHz flux ratio.
    \item Columns 24--25: Logarithm of the blazar SED synchrotron peak computed using the Blazar Synchrotron Tool, BlaST \citep{glauch22}.
\end{itemize}

\end{document}